\documentclass[twocolumn]{aastex62}
\usepackage[]{threeparttable}

\graphicspath{{./}{figures/}}

\shorttitle{{\it HST} observations of MADCASH dwarfs}
\shortauthors{Carlin et al.}

\begin{document}

\title{{\it Hubble Space Telescope} Observations of Two Faint Dwarf Satellites of Nearby LMC Analogs from MADCASH\footnote{Based in part on data collected at Subaru Telescope, which is operated by the National Astronomical Observatory of Japan.}}

\correspondingauthor{Jeffrey L. Carlin}
\email{jcarlin@lsst.org, jeffreylcarlin@gmail.com}

\author[0000-0002-3936-9628]{Jeffrey L. Carlin}
\affil{NSF's NOIRLab/ Rubin Observatory Project Office, 950 North Cherry Avenue, Tucson, AZ 85719, USA} 

\author[0000-0001-9649-4815]{Bur\c{c}\.{i}n Mutlu-Pakd\.{i}l}
\affil{Kavli Institute for Cosmological Physics, University of Chicago, Chicago, IL 60637, USA}
\affil{Department of Astronomy and Astrophysics, University of Chicago, Chicago IL 60637, USA}
\affil{Department of Astronomy/Steward Observatory, 933 North Cherry Avenue, Rm. N204, Tucson, AZ 85721-0065, USA}

\author[0000-0002-1763-4128]{Denija Crnojevi\'{c}}
\affil{University of Tampa, 401 West Kennedy Boulevard, Tampa, FL 33606, USA}

\author[0000-0001-9061-1697]{Christopher T. Garling}
\affil{CCAPP and Department of Astronomy, The Ohio State University, Columbus, OH 43210, USA}

\author[0000-0001-8855-3635]{Ananthan Karunakaran}
\affil{Department of Physics, Engineering Physics and Astronomy, Queen's University, Kingston, Ontario, Canada, K7L 3N6}

\author[0000-0002-8040-6785]{Annika H. G. Peter}
\affil{CCAPP, Department of Physics, and Department of Astronomy, The Ohio State University, Columbus, OH 43210, USA}

\author{Erik Tollerud}
\affil{Space Telescope Science Institute, 3700 San Martin Drive, Baltimore, MD 21218, USA}

\author{Duncan A. Forbes}
\affil{Centre for Astrophysics and Supercomputing, Swinburne University, Hawthorn VIC 3122, Australia}

\author{Jonathan R. Hargis}
\affil{Space Telescope Science Institute, 3700 San Martin Drive, Baltimore, MD 21218, USA}

\author[0000-0002-5049-4390]{Sungsoon Lim}
\affil{University of Tampa, 401 West Kennedy Boulevard, Tampa, FL 33606, USA}

\author{Aaron J. Romanowsky}
\affil{Department of Physics \& Astronomy, San Jos\'e State University, One Washington Square, San Jose, CA 95192, USA}
\affil{University of California Observatories, 1156 High Street, Santa Cruz, CA 95064, USA}

\author{David J. Sand}
\affil{Department of Astronomy/Steward Observatory, 933 North Cherry Avenue, Rm. N204, Tucson, AZ 85721-0065, USA}

\author{Kristine Spekkens}
\affil{Department of Physics, Engineering Physics and Astronomy, Queen's University, Kingston, Ontario, Canada, K7L 3N6}
\affil{Department of Physics and Space Science, Royal Military College of Canada, P.O. Box 17000, Station Forces, Kingston, ON K7K 7B4, Canada}

\author{Jay Strader}
\affil{Department of Physics and Astronomy, Michigan State University,East Lansing, MI 48824, USA}

\begin{abstract}
We present a deep {\it Hubble Space Telescope} ({\it HST}) imaging study of two dwarf galaxies in the halos of Local Volume Large Magellanic Cloud (LMC) analogs. These dwarfs were discovered as part of our Subaru$+$Hyper Suprime-Cam MADCASH survey: MADCASH-1, which is a satellite of NGC~2403 ($D\sim3.2$~Mpc), and MADCASH-2, a previously unknown dwarf galaxy near NGC~4214 ($D\sim3$~Mpc). Our {\it HST} data reach $>3.5$~mag below the tip of the red giant branch (TRGB) of each dwarf, allowing us to derive their structural parameters and assess their stellar populations. We measure TRGB distances ($D_{\text{MADCASH-1}}=3.41^{+0.24}_{-0.23}$~Mpc, $D_{\text{MADCASH-2}}=3.00^{+0.13}_{-0.15}$~Mpc), and confirm the dwarfs' associations with their host galaxies. MADCASH-1 is a predominantly old, metal-poor stellar system (age $\sim$13.5~Gyr, [M/H] $\sim -2.0$), similar to many Local Group dwarfs. Modelling of MADCASH-2's CMD suggests that it contains mostly ancient, metal-poor stars (age $\sim13.5$ Gyr, [M/H] $\sim -2.0$), but that $\sim10\%$ of its stellar mass was formed 1.1--1.5 Gyr ago, and $\sim1\%$ was formed 400--500 Myr ago. Given its recent star formation, we search MADCASH-2 for neutral hydrogen using the Green Bank Telescope, but find no emission and estimate an upper limit on the H\textsc{I} mass of $< 4.8 \times 10^4 \ M_{\odot}$. These are the faintest dwarf satellites known around host galaxies of LMC mass outside the Local Group ($M_{V,\text{MADCASH-1}}=-7.81\pm0.18$, $M_{V,\text{MADCASH-2}}=-9.15\pm0.12$),
and one of them shows signs of recent environmental quenching by its host.  Once the MADCASH survey for faint dwarf satellites is complete, our census will enable us to test CDM predictions for hierarchical structure formation, and discover the physical mechanisms by which low-mass hosts influence the evolution of their satellites.

\end{abstract}

\section{Introduction} \label{sec:intro}
Large dark matter halos grow and evolve via the merging of smaller subhalos with their more massive host \citep[e.g.,][]{klypin1999, moore1999}. In these mergers the dark matter and baryons of the accreted satellite are assimilated into the more massive system. Cosmological simulations of structure formation \citep[e.g.,][]{Springel2008a} generally predict that the hierarchy of dark matter substructure is essentially scale-free -- the number of subhalos scales with the total mass of the host. However, this scale invariance does not carry over to the baryonic component of galaxies.  Due to a combination of environmental (e.g., reionization quenching and ram pressure stripping) and self-regulatory (e.g, AGN and supernova feedback) effects, the mapping between subhalo mass and stellar mass is non-linear. Therefore, the detailed properties of the baryonic components of subhalos -- dwarf satellite galaxies -- cannot be easily inferred from the statistics of dark matter substructure predicted by cosmological models. The luminosity functions (LFs), spatial distributions, metallicities, and star formation histories (SFHs) of dwarf satellite systems depend not only on the physics of the dark matter that provides the dense ``seeds'' in which these galaxies form, but also on the environment in which these satellites form and evolve \citep[see, e.g., reviews by][]{Bullock2017, Wechsler2018}. In order to use dwarf satellite populations to constrain the underlying dark matter and baryonic physics, it is important to compile complete samples of dwarfs around hosts covering a wide range of mass and residing in a variety of environments. 

In the past two decades, deep, large sky area digital imaging surveys have increased the number of known dwarf galaxies around the Milky Way \citep[MW; e.g.,][]{Willman2005, Willman2005a,Belokurov2006a,Belokurov2006b,Belokurov2007b,Belokurov2008,Belokurov2010,Zucker2006,Zucker2006a,Bechtol2015,Drlica-Wagner2015,Drlica-Wagner2016,Kim2015,Laevens2015,Homma2016,Homma2018,Homma2019,Torrealba2016,Torrealba2018,Mau2019,Cerny2020} and our nearest massive neighbor M31 \citep[e.g.,][]{Martin2016a, McConnachie2018} from roughly a dozen to nearly 100 satellites. These include the ultra-faint dwarfs (UFDs), a class of low-luminosity galaxies that are dark matter dominated systems typically consisting only of ancient, extremely metal-poor stellar populations \citep[see review by][]{Simon2019}. The more luminous ($L_{\text{V}} \gtrsim 10^5 L_\odot$) systems are typically referred to as the ``classical'' dwarf spheroidals (dSphs). 

Large-aperture telescopes equipped with wide-field imaging cameras are now making it possible to search for satellites of massive, MW-like host galaxies in the Local Volume (LV; $D\lesssim11$~Mpc). Large, nearby galaxies for which a satellite census has been performed include Centaurus A \citep{Taylor2018,Crnojevic2019,Muller2019}, M~81 \citep{Chiboucas2013}, M~94 \citep{Smercina2018}, M~101 \citep{Merritt2014,Danieli2017,Bennet2019,Bennet2020}, and NGC~253 \citep{Sand2014, Romanowsky2016,Toloba2016a}, as well as more distant LV systems \citep[e.g.,][]{Carlsten2020,Davis2020}. Dwarfs satellites discovered around these systems  are essential for making comparisons to predictions from simulations of massive galaxies \citep[e.g.,][]{Benson2002,Zolotov2012,Wetzel2016,Jethwa2018,Bose2018,Kim2018,Nadler2019,garrison-kimmel2019b,samuel2020}. However, for lower mass hosts, halo-to-halo scatter in number of dwarfs and/or their LF, the stellar mass-halo mass relation \citep[SMHM; e.g.,][]{Behroozi2013,Moster2013,Brook2014,Garrison-Kimmel2014,Garrison-Kimmel2017,Munshi2018}, and effects such as reionization, ram pressure, tides, and infall time may be relatively more important in shaping the physical properties of dwarf satellites \citep[e.g.,][]{Dooley2017b}. 

To explore satellite populations around lower mass hosts, we have undertaken the MADCASH (Magellanic Analog Dwarf Companions and Stellar Halos) project, a deep, ground-based imaging survey in which we are systematically mapping the resolved stellar halos of nearby ($D\lesssim4$~Mpc) Magellanic Cloud (MC) analogs (i.e., galaxies with stellar masses between $\sim0.1~M_{\text{star,SMC}}$ to $\sim3~M_{\text{star, LMC}}$). The spatial clustering of many of the $\sim$20 UFDs discovered in the Dark Energy Survey (DES) and other southern sky surveys provides a tantalizing hint that the LMC fell into the MW with its own satellite system \citep[e.g.,][]{Bechtol2015, Drlica-Wagner2015, Koposov2015, Jethwa2016, Sales2017, Kallivayalil2018, Erkal2020, Nadler2020}. With the MADCASH survey, we are measuring the satellite populations of \textit{field} LMC-mass galaxies, in order to increase the number of MC-mass systems with well-characterized dwarf LFs, while also avoiding the complicating influence of the MW's gravitational potential on interpretation of LMC satellites. \citet{Dooley2017b} used the dark-matter-only Caterpillar cosmological simulations in concert with abundance matching methods to predict that a system with the stellar mass of the Large Magellanic Cloud ($M_{\text{star, LMC}}\sim 2$--$3\times10^9 M_{\odot}$, or $\sim1/20$ the stellar mass of the MW, e.g. \citealt{Kim1998,Harris2009}) should host $\sim2-5$ satellites with $L > 10^5 L_{\odot}$, and as many as 15 dwarf satellites when including the UFDs \citep[see also][]{Jahn2019,Nadler2020}.

In this work we present {\it HST} observations of the first two dwarf galaxies discovered by the MADCASH survey. 
To confirm their nature as dwarf satellites of MC-mass hosts, and to extract reliable structural and stellar population properties of these faint dwarfs, we require the superior resolution of {\it HST}. MADCASH-1 (aka MADCASH J074238+652501-dw; \citealt{Carlin2016}) is the first dwarf we reported from our survey. It is located $\sim35$~kpc in projection from LMC analog NGC~2403, and, with a luminosity near the dividing line between UFDs and
classical dwarfs, is the faintest known dwarf satellite of a $\sim$MC-mass host. The second dwarf we explore in this work -- 
called MADCASH-2 -- has not been previously announced. It is a likely satellite of NGC~4214, but is too compact in our ground-based Subaru discovery imaging to reliably extract its properties. With the {\it HST} observations reported here, we confirm its nature as a dwarf galaxy at the distance of NGC~4214, and explore its luminosity, structural parameters, and stellar populations.

\section{Ground-based discovery of the first two MADCASH dwarfs} \label{sec:discovery}

Both of the dwarf galaxies discussed in this work were discovered by visual examination of deep images from Hyper Suprime-Cam (HSC; \citealt{Miyazaki2012}) on the 8.2m Subaru telescope. The observations and discovery of MADCASH-1 were detailed in \citet{Carlin2016}; MADCASH-2 was subsequently identified in the imaging data around NGC~4214. These data consist of 12x300s = 3600s sequences of images in $g$-band (``HSC-G'' in Subaru parlance) and 12x120s = 1440s in $i$-band (``HSC-I''). 
Offsets and rotational dithers were applied to each exposure to facilitate cosmic-ray removal and to fill chip gaps. Raw data were processed through all steps including source detection and extraction using the development version of the LSST pipeline (see, e.g., \citealt{Bosch2018, Aihara2018a, Aihara2018b, Aihara2019} for details on the processing as applied to HSC). The candidate dwarf MADCASH-2 was discovered in a visual search of the resulting images, after which we obtained follow-up {\it HST} observations for both dwarfs.

\subsection{MADCASH-1 - NGC 2403 satellite}

``MADCASH-1'' is the shorthand we will use for this dwarf galaxy, which is officially named MADCASH J074238+652501-dw \citep{Carlin2016}. It was found in a visual search of Subaru+HSC images of the region around NGC~2403. It is partially resolved in the Subaru images, but there are only a few stars near its RGB. 

\begin{figure*}[!t]
\includegraphics[width=0.47\textwidth, trim=2.5in 1.0in 2.5in 1.0in, clip]{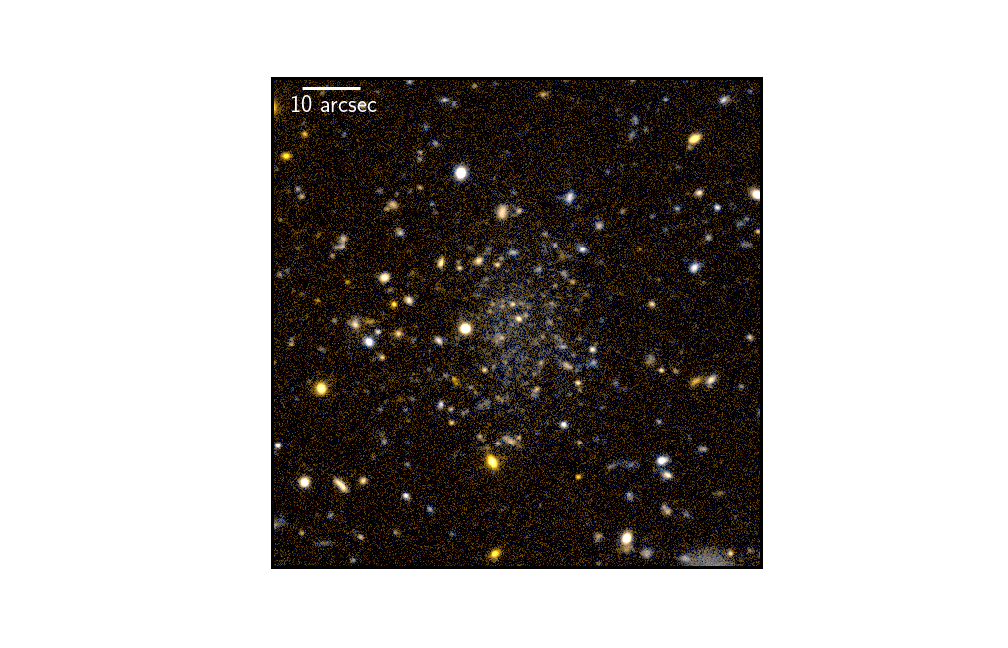}
\includegraphics[width=0.525\textwidth, trim=0.5in 0.0in 1.0in 1.0in, clip]{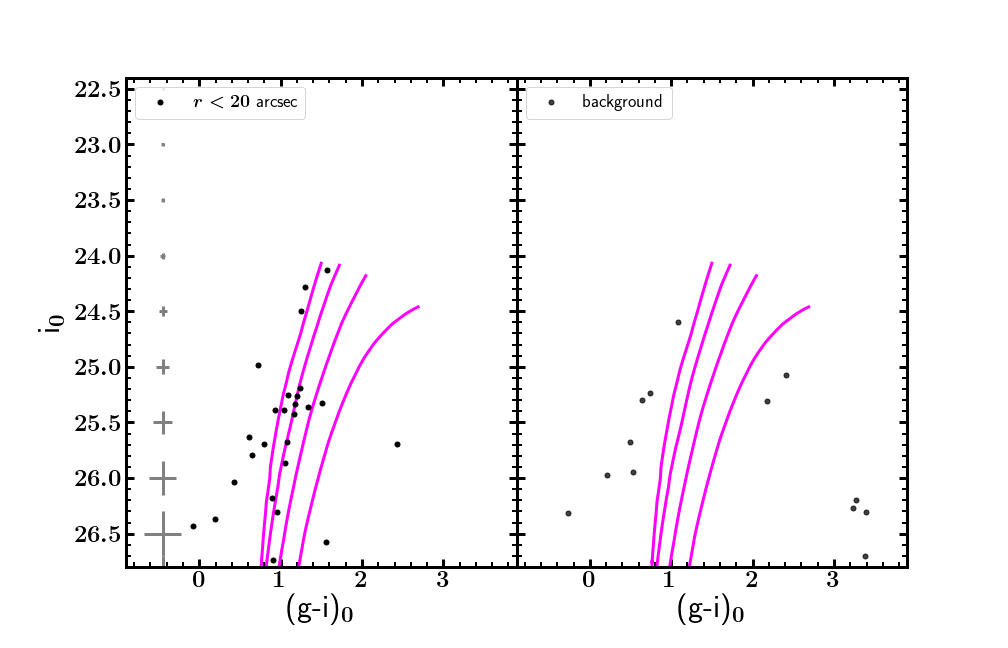}
\includegraphics[width=0.47\textwidth, trim=2.5in 1.0in 2.5in 1.0in, clip]{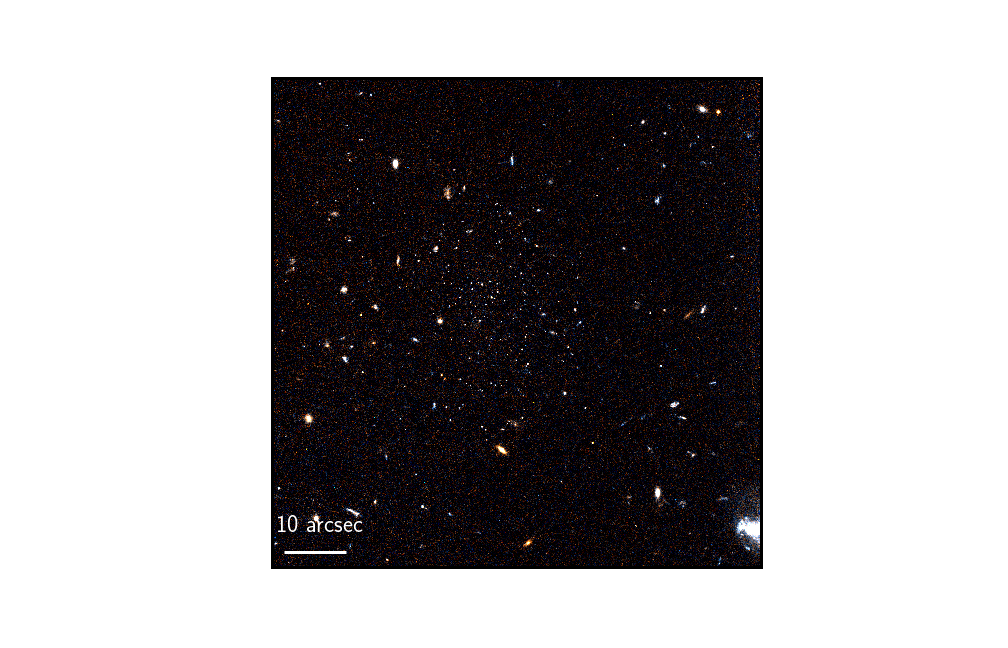}
\includegraphics[width=0.525\textwidth, trim=0.5in 0.0in 1.0in 1.0in, clip]{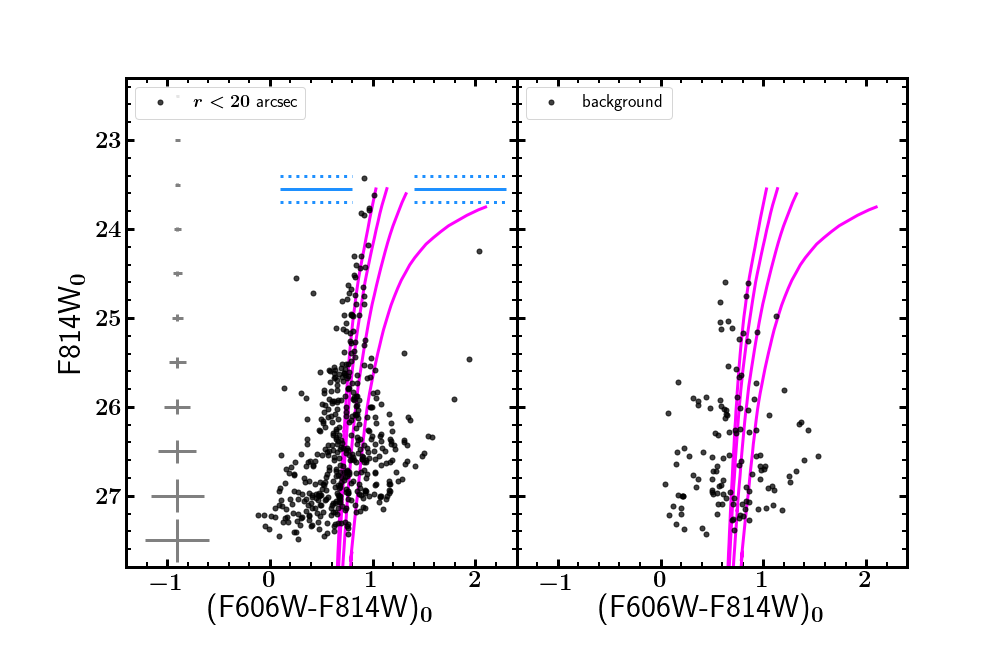}
\caption{{\it Upper left:} Subaru color image of MADCASH-1, created from a combination of the $g$- and $i$-band images. {\it Upper right:} Color-magnitude diagram of stars near MADCASH-1 (left panel). The right panel shows a random background region of the same size. Both panels include 13.5-Gyr PARSEC isochrones \citep{Aringer2009,Bressan2012,Chen2014,Marigo2017} with metallicities (from left to right) of [M/H]$ = -2.0, -1.5, -1.0$, and $-0.5$, shifted to an assumed distance of 3.41~Mpc (see Sec.~\ref{subsec:distances}). Crosses at the far left show the median magnitude and color errors in 0.5-mag bins. A red giant branch corresponding to MADCASH-1 is visible in the left panel, with no corresponding feature in the background region. However, because of crowding, the photometry does not produce a well-defined RGB. {\it Lower left:} ACS color image of MADCASH-1, created from a combination of the $F606W$- and $F814W$-band images. {\it Lower right:} Color-magnitude diagram of stars near MADCASH-1 (left panel). The right panel shows a random background region of the same size from the other side of the ACS field. Both CMDs include the same isochrones as in the upper panels, but for the {\it HST}+ACS bands. A red giant branch corresponding to MADCASH-1 is clearly visible in the left panel, with no corresponding feature in the background region. Horizontal blue lines mark the location of the TRGB (see Sec.~\ref{subsec:distances}), with dotted lines signifying the uncertainty on the TRGB magnitude.}
\label{fig:madcash1_subaru_acs}
\end{figure*}

In Figure~\ref{fig:madcash1_subaru_acs}, we compare the Subaru+HSC image and color-magnitude diagram (CMD) with those from our {\it HST} observations. The superior resolution of {\it HST} clearly resolves MADCASH-1 into individual stars, yielding a well-defined RGB of an old, metal-poor population in the CMD. We use the {\it HST} data to confirm that MADCASH-1 is a dwarf galaxy at roughly the same distance as NGC~2403, and to measure its structural parameters and assess its stellar populations.

\subsection{MADCASH-2 - NGC 4214 satellite?}

\begin{figure*}[!t]
\begin{center}
\includegraphics[width=0.8\textwidth, trim=0in 0.0in 0in 0.0in, clip]{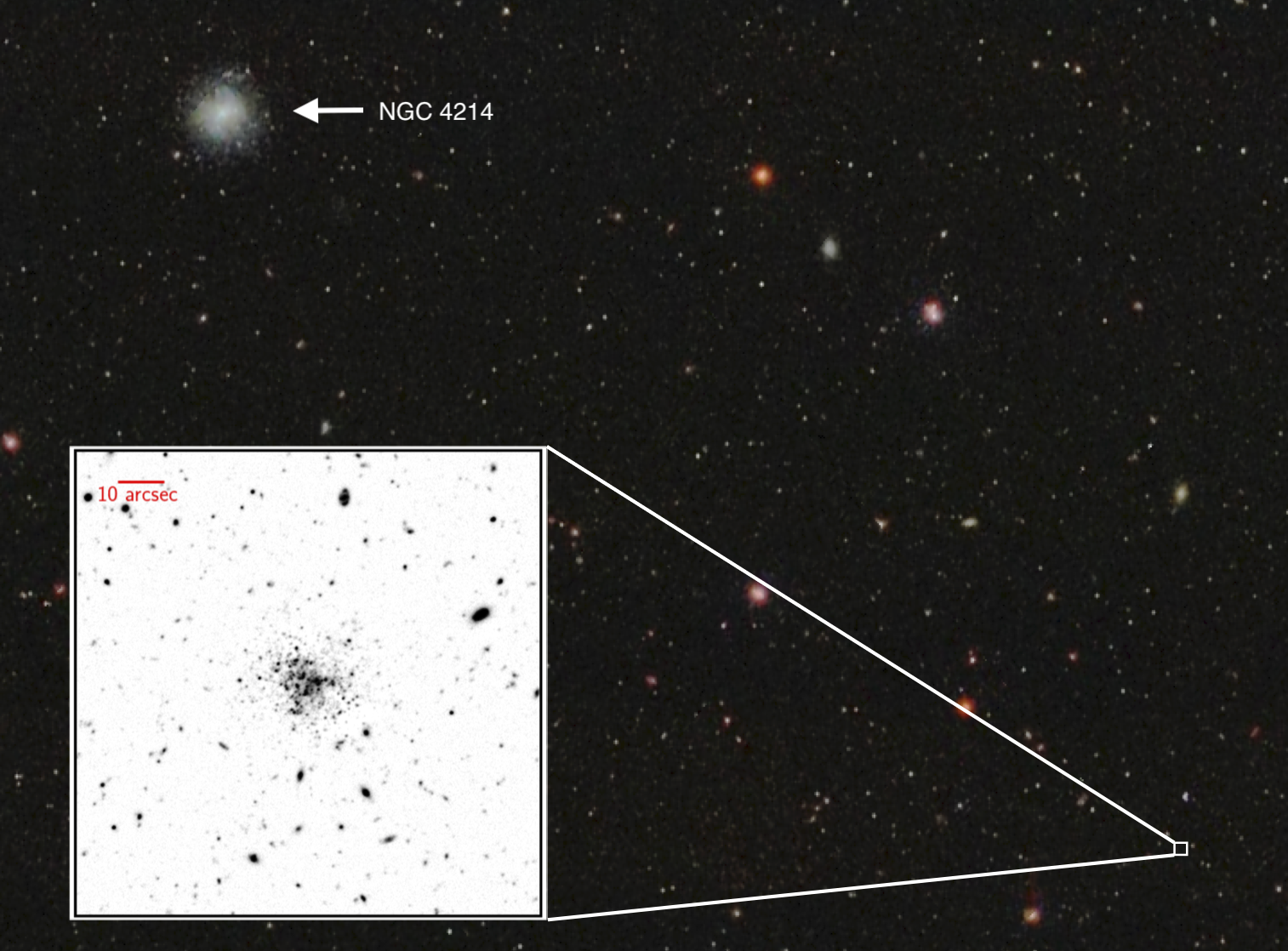}
\caption{SDSS DR9 color image (from Aladdin: \citealt{Aladdin2000,Aladdin2014}) of the region to the southwest of NGC~4214; north is up, and east to the left. NGC~4214 is the large galaxy at the upper left. The inset shows a Subaru+HSC $g$-band image of MADCASH-2, which lies  $\sim1.4^\circ$ ($\sim70$~kpc at the distance of NGC~4214) in projection from NGC~4214.}
\label{fig:madcash2_subaru_sdss}
\end{center}
\end{figure*}

\begin{figure*}[!t]
\includegraphics[width=0.47\textwidth, trim=2.5in 1.0in 2.5in 1.0in, clip]{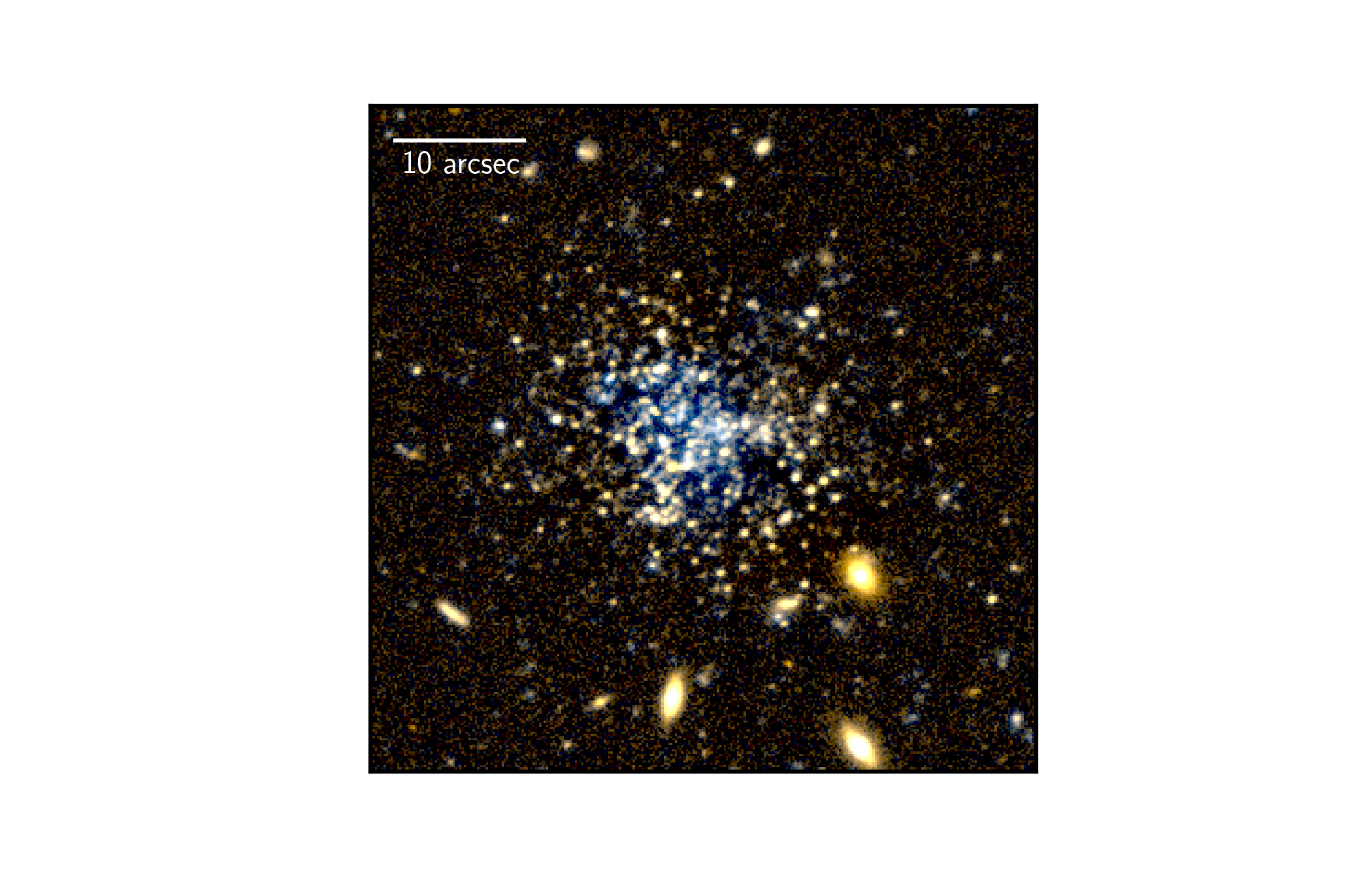}
\includegraphics[width=0.525\textwidth, trim=0.5in 0.0in 1.0in 1.0in, clip]{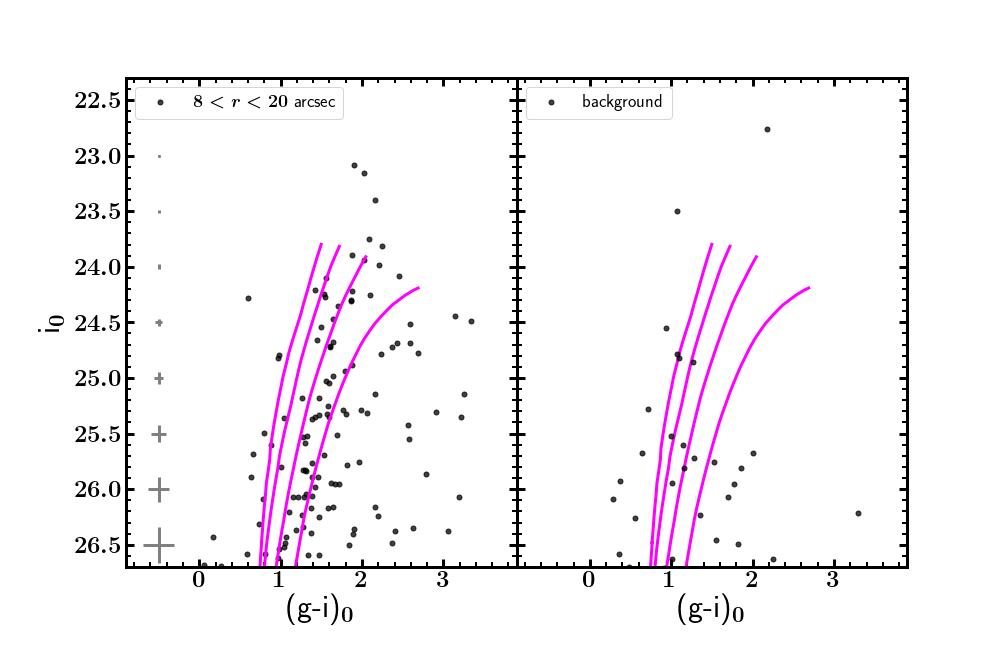}
\includegraphics[width=0.47\textwidth, trim=2.5in 1.0in 2.5in 1.0in, clip]{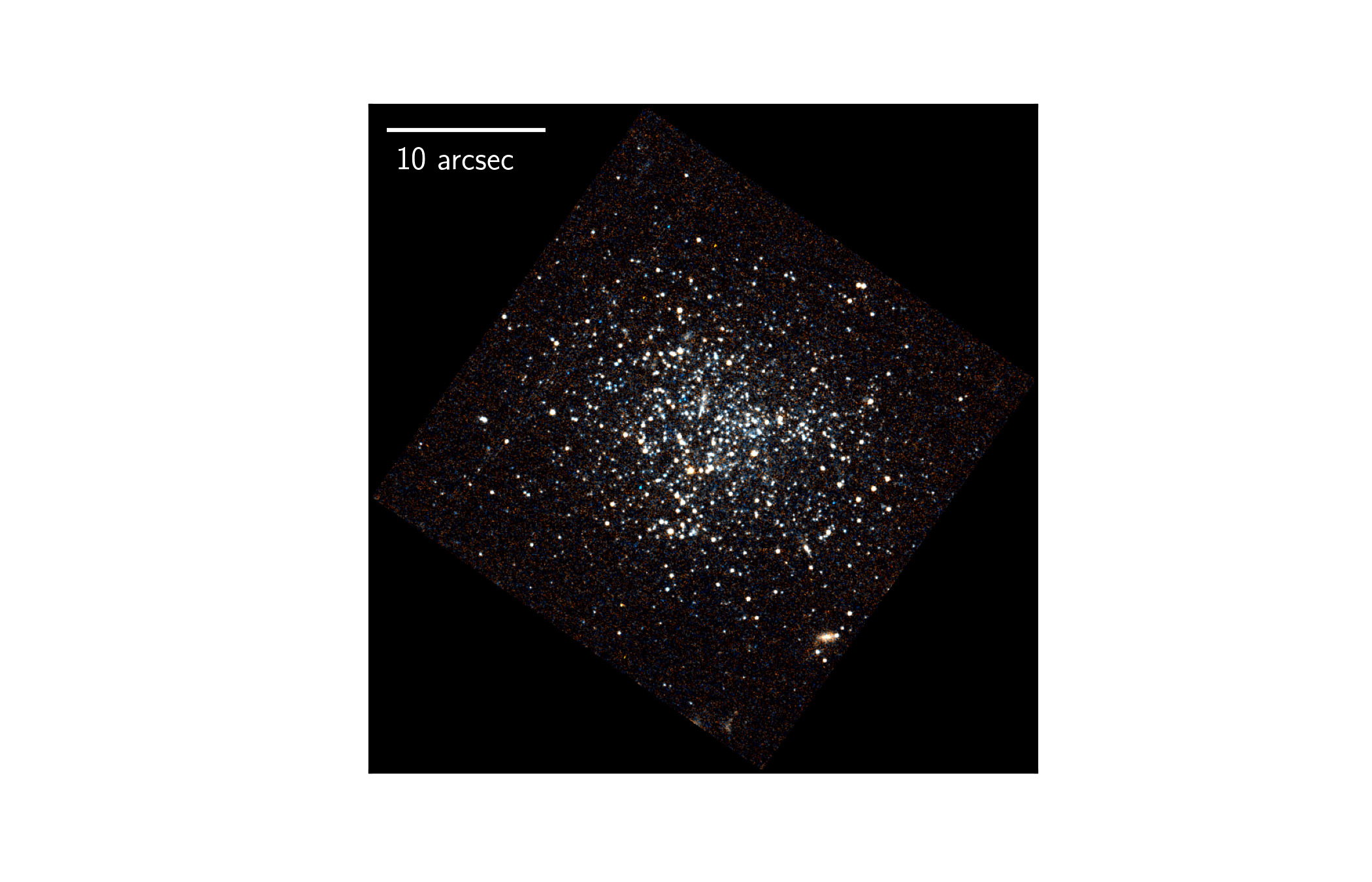}
\includegraphics[width=0.525\textwidth, trim=0.5in 0.0in 1.0in 1.0in, clip]{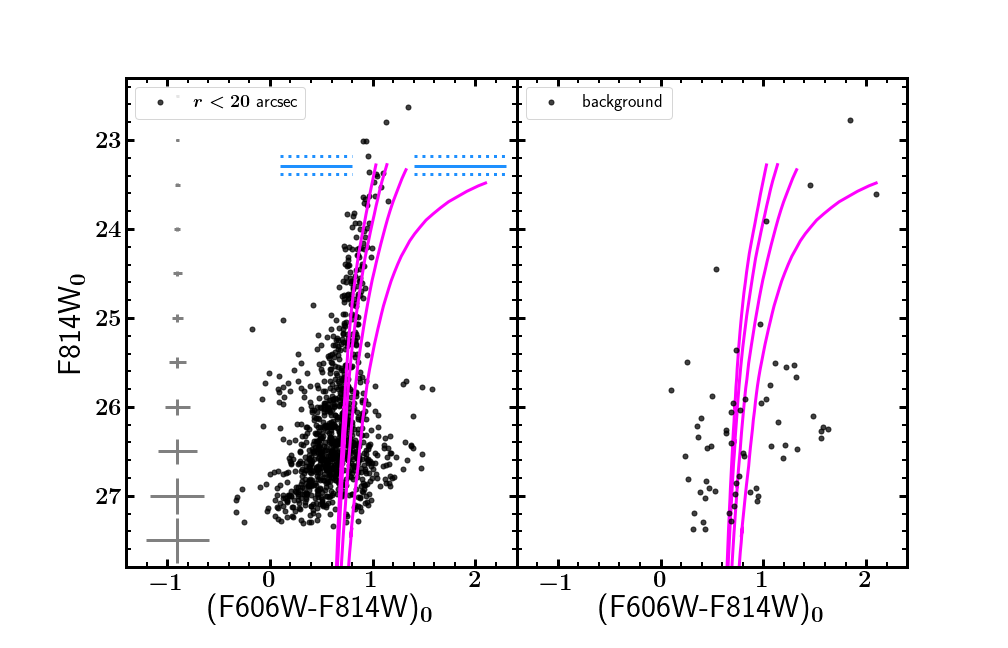}
\caption{{\it Upper left:} Subaru color image of MADCASH-2, created from a combination of the $g$- and $i$-band images. {\it Upper right:} Color-magnitude diagram of stars near MADCASH-2 (left panel), excluding the innermost region where crowding renders stars unmeasurable. The right panel shows a random background region of the same size. Both panels include 13.5-Gyr 
PARSEC isochrones with metallicities (from left to right) of [M/H]$ = -2.0, -1.5, -1.0$, and $-0.5$, shifted to an assumed distance of 3.00~Mpc (see Sec.~\ref{subsec:distances}). Crosses at the far left show the median magnitude and color errors in 0.5-mag bins. A red giant branch corresponding to MADCASH-2 is clearly visible in the left panel, with no corresponding feature in the background region. However, because of crowding, the photometry does not produce a well-defined RGB. {\it Lower left:} ACS color image of MADCASH-2, created from a combination of the $F606W$- and $F814W$-band images. {\it Lower right:} Color-magnitude diagram of stars near MADCASH-2 (left panel). The right panel shows a random background region of the same size. Both CMDs include the same isochrones as in the upper panels, but for the {\it HST}+ACS bands. A red giant branch corresponding to MADCASH-2 is clearly visible in the left panel, with no corresponding feature in the background region. Horizontal blue lines mark the location of the TRGB (see Sec.~\ref{subsec:distances}), with dotted lines signifying the uncertainty on the TRGB magnitude.}
\label{fig:madcash2_subaru_acs}
\end{figure*}

The second dwarf discovered as part of our survey is tentatively called ``MADCASH-2'' for simplicity, but is officially named MADCASH~J121007+352635-dw (following IAU naming conventions). It was first seen by eye in examination of deep Subaru+HSC images around NGC~4214. In Figure~\ref{fig:madcash2_subaru_sdss}, we show the location of MADCASH-2 relative to NGC~4214, along with an inset image of the candidate dwarf from the Subaru/HSC data. The object is compact, with very few RGB stars resolved photometrically, so that we are unable to determine its properties or its association to NGC 4214 from the ground-based data. In Figure~\ref{fig:madcash2_subaru_acs}, we compare the Subaru and {\it HST} images and CMDs. MADCASH-2 easily resolves into its individual stars in the {\it HST} image, yielding a deep, detailed CMD consistent with its being a classical dwarf spheroidal (dSph) at roughly the distance of NGC~4214.

\begin{figure}[!t]
\includegraphics[width=0.9\columnwidth,trim=0.5in 0.0in 0.5in 0.5in, clip]{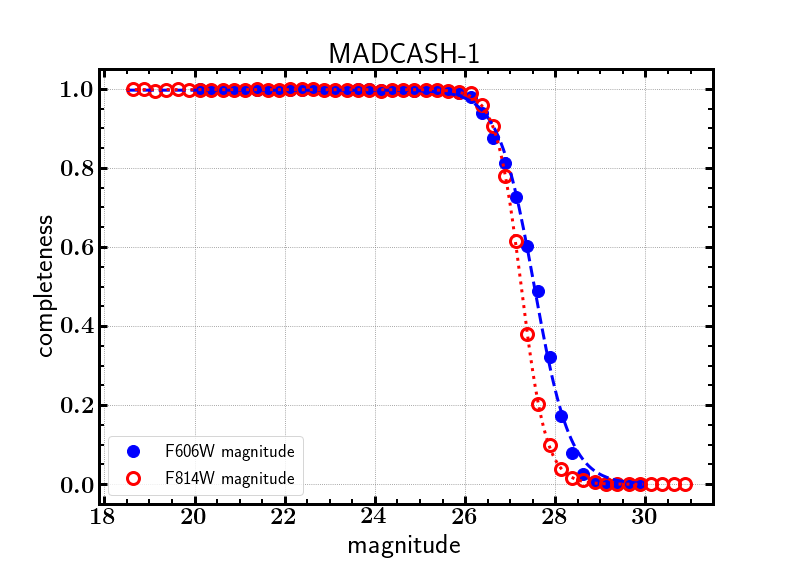}\\
\includegraphics[width=0.9\columnwidth,trim=0.5in 0.0in 0.5in 0.5in, clip]{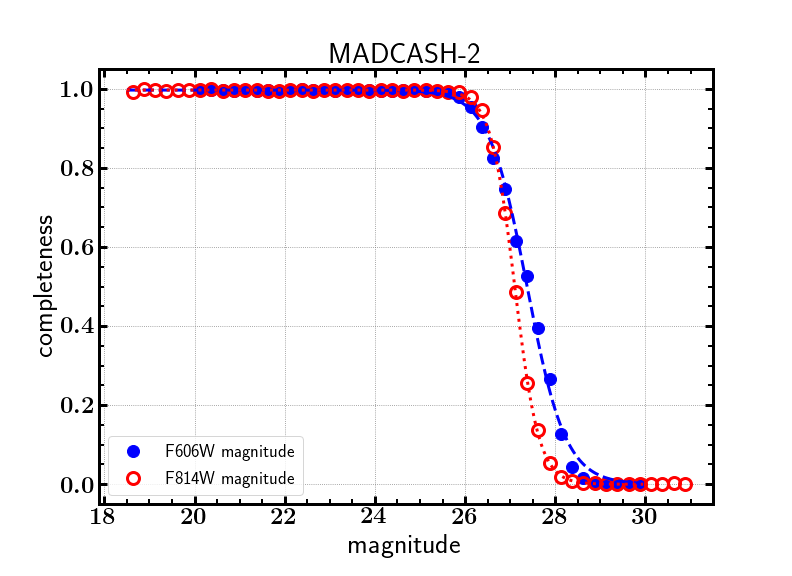}
\caption{Completeness as a function of magnitude for both fields, as determined from artificial star tests. The overlaid curves are fits to the data 
using Equation~7 from \citet{Martin2016a}. From the fits, we estimate that the MADCASH-1 data are 50\% complete at F606W = 27.54 and F814W = 27.25, while the MADCASH-2 data are 50\% complete at F606W = 27.37 and F814W = 27.10 (see Table~\ref{tab:obslog}).
}
\label{fig:completeness}
\end{figure}

\begin{figure*}[!t]
\includegraphics[width=0.425\textwidth, trim=0.0in 0.0in 0.0in 0.0in, clip]{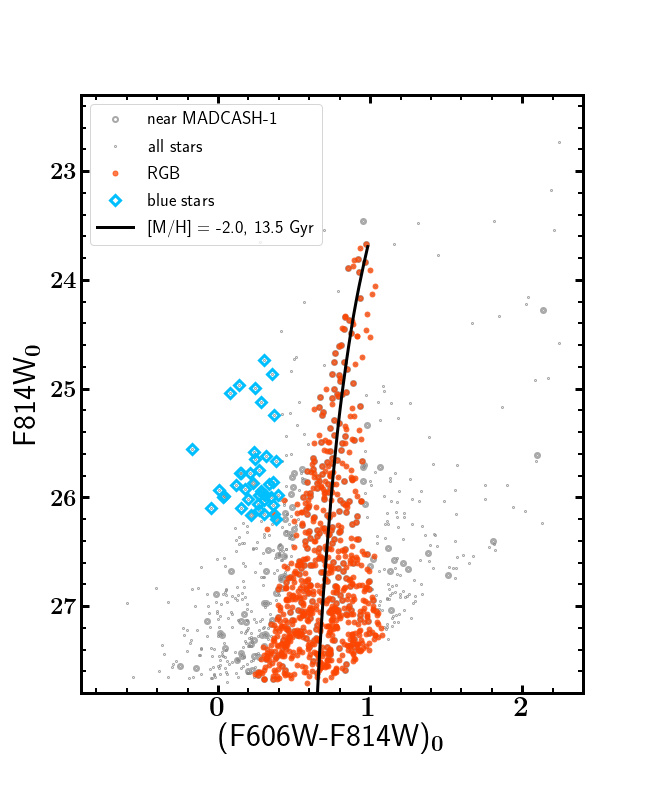}
\includegraphics[width=0.57\textwidth, trim=0.0in 0.0in 1.0in 1.0in, clip]{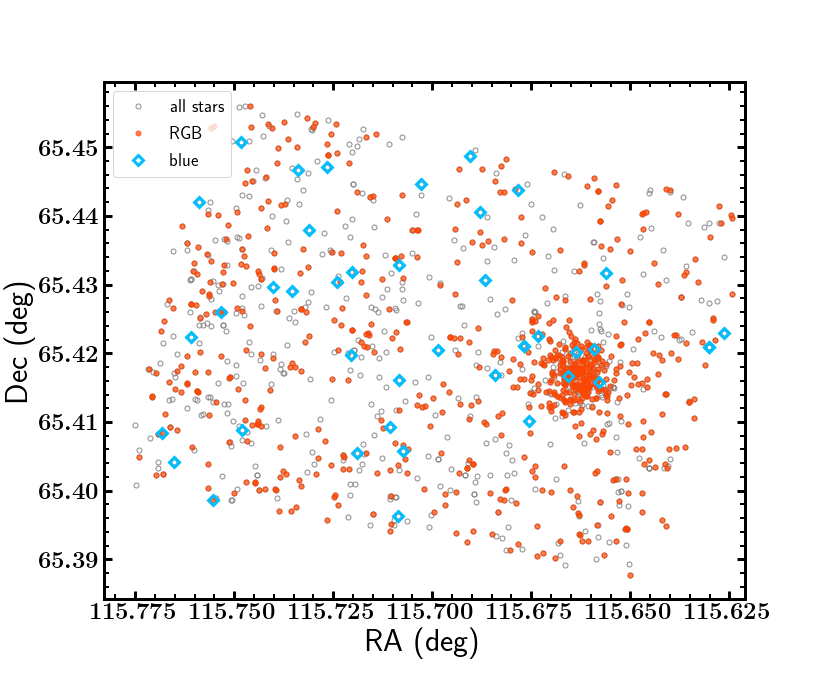}
\caption{Spatial distribution of MADCASH~1 stars from the {\it HST} imaging. {\it Left:} CMD showing the sub-populations we have selected from the {\it HST} imaging of MADCASH-1. Blue stars (selected in a CMD region where intermediate-age RGB stars would appear in MADCASH-1) are blue diamonds, and the RGB is shown as red points. {\it Right:} Spatial distribution of the sub-populations from the left panel (symbols are the same in both panels). MADCASH-1 is evident as a large overdensity near the right side of the {\it HST} field of view (it was deliberately placed to one side of the FOV to avoid chip gaps). The blue objects are not concentrated near the center of MADCASH-1, and are thus unlikely to be associated with the dwarf. }
\label{fig:madcash1_subpops}
\end{figure*}

\begin{figure*}[!t]
\includegraphics[width=0.425\textwidth, trim=0.0in 0.0in 0.0in 0.0in, clip]{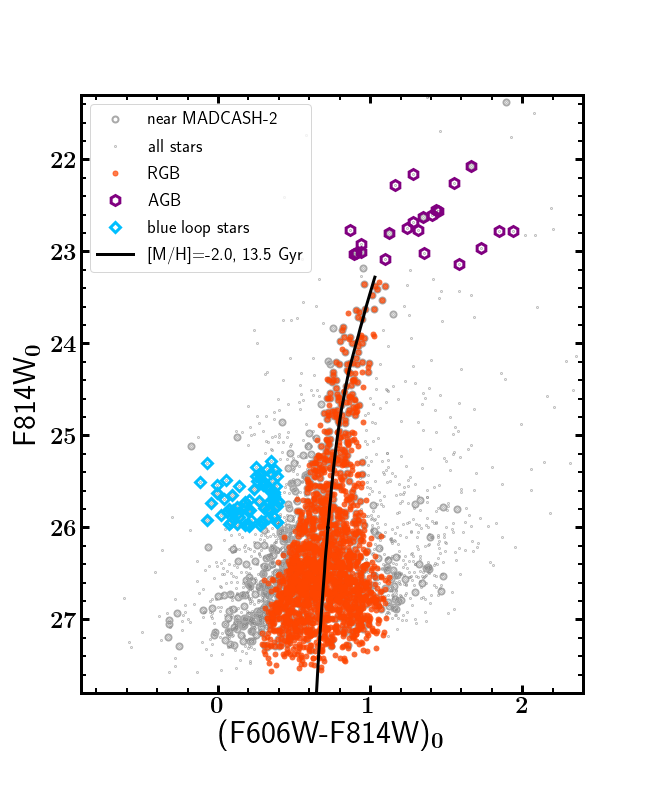}
\includegraphics[width=0.57\textwidth, trim=0.0in 0.0in 1.0in 1.0in, clip]{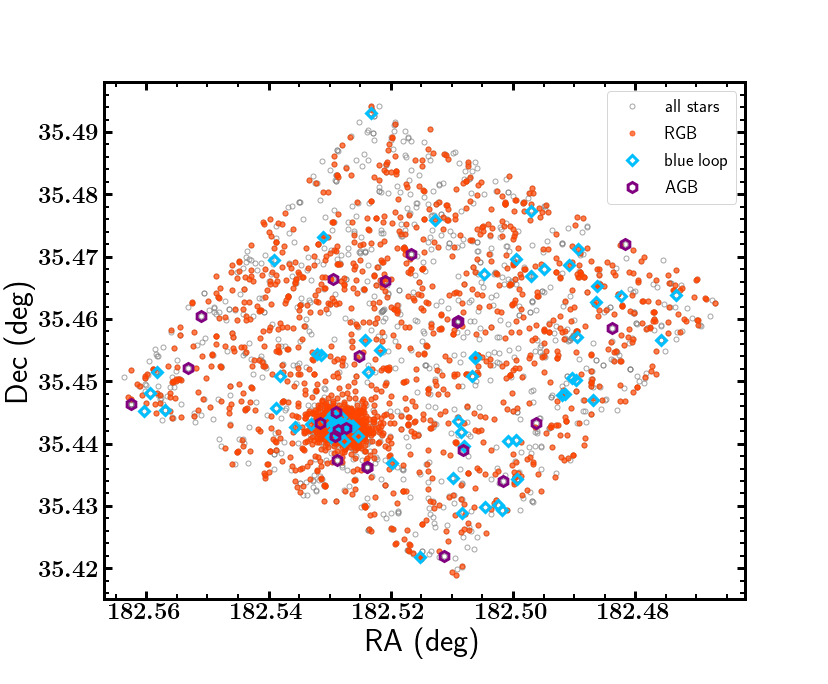}
\caption{Spatial distribution of MADCASH~2 stars from the {\it HST} imaging. {\it Left:} CMD showing the sub-populations we have selected from {\it HST} imaging of MADCASH-2. Purple hexagons are candidate AGB stars, blue stars (including possible young populations in MADCASH-2) are blue diamonds, and the RGB is shown as red points. {\it Right:} Spatial distribution of the sub-populations from the left panel (symbols are the same in both panels). MADCASH-2 is evident as a large overdensity near the bottom of the {\it HST} field of view. There is also a clear concentration of blue and AGB stars at the same position, confirming that these stars are indeed associated with MADCASH-2. }
\label{fig:madcash2_subpops}
\end{figure*}

\begin{figure*}[!t]
\includegraphics[width=1.0\textwidth, trim=0.0in 0.0in 0.0in 0in, clip]{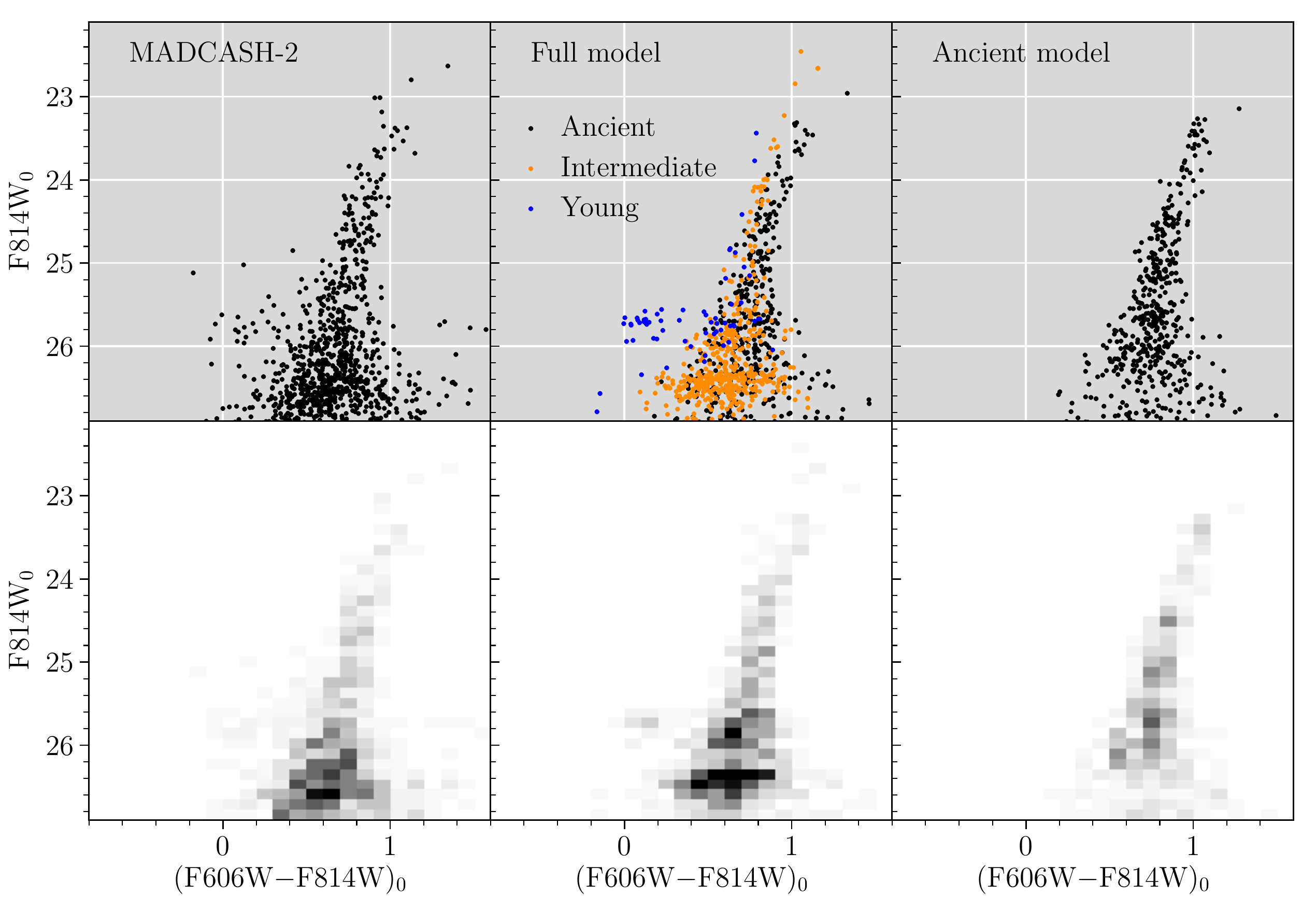}
\caption{\textit{Left column:} CMD (upper) and Hess diagram (lower panel) of stars within $\sim2~r_{\rm h}$ of MADCASH-2. \textit{Right column:} Simulated ancient (13.5~Gyr), metal-poor ([M/H]$=-2.0$) stellar population containing 90\% of the stellar mass of MADCASH-2, at the distance of MADCASH-2, and convolved with the completeness and photometric errors of our {\it HST} data as a function of magnitude. In both rows, it is obvious that this ancient stellar population alone cannot account for the morphology of the MADCASH-2 CMD. We thus include younger stellar populations to account for (a) the ``blue loop'' stars at $25.5 \lesssim$ F814W $\lesssim $26.0, and (b) the excess of stars at F814W$_0$ $>$ 26 in the MADCASH-2 CMD relative to the simulated ancient population. The \textit{center panels} show our solution that best matches the CMD morphology of MADCASH-2. In addition to the ancient stellar population, we include $\sim10\%$ of the stellar mass in an intermediate-age (1.1--1.5~Gyr old), metal-poor ([M/H]$=-1.5$) population (orange points in the CMD), and a young ($\sim400$~Myr) population accounting for $\sim1\%$ of the stellar mass. The youngest population (in blue in the top center panel) reproduces the very blue stars seen in the CMD; we note that slightly older populations do not extend far enough blueward of the RGB to match this population. Likewise, the apparent paucity of stars at F814W$_0 > 26$ is well explained by a blue loop population, but this time of more intermediate age. Both young populations also contribute a little bit to the broadening of the upper RGB compared to what a pure ancient population would predict. This composite stellar population reproduces the general properties of the MADCASH-2 CMD, confirming that this dSph contains a small fraction of young stars, and a significant portion of its stars formed in the past $\sim2$~Gyr. 
}
\label{fig:madcash2_cmd_sims}
\end{figure*}

\section{Observations and Data Reduction} \label{sec:data}

We obtained {\it HST} optical observations of MADCASH-1 and MADCASH-2, using the F606W and F814W filters on the Advanced Camera for Surveys (ACS; HST-GO-15228; PI: J. Carlin) with a single orbit per filter per targeted galaxy (see observation summary in Table~\ref{tab:obslog}).  These observations allow us to reach $\gtrsim 3.5$ mag below the tip of the Red Giant Branch (TRGB, i.e., down to $I\gtrsim27.0$) with a signal-to-noise ratio of $\sim5$. A standard 4-point dither pattern was used to minimize the effect of detector defects and optimally sample the PSF. 

We performed point-spread function photometry on the flat-fielded (FLT) images using the latest version (2.0) of DOLPHOT \citep{Dolphin2002}, an updated version of HSTPHOT \citep{Dolphin2000}, largely using the recommended prescriptions. Photometry was performed on the individual FLT images using the drizzled DRZ images as an astrometric reference frame. The photometry was then culled with the following criteria to select well-measured stars: ($sharpness_{\rm F606W}+sharpness_{\rm F814W})^2 < 0.075$, ($crowd_{\rm F606W}+crowd_{\rm F814W}) < 1.0$, signal-to-noise ratio $> 4$, and object-type $\leq 2$ in each filter. We corrected for Milky Way extinction on a star-by-star basis using the \citet{Schlegel1998} reddening maps with the coefficients from \citet{Schlafly2011}. Tables~\ref{tab:Catalog1}-\ref{tab:Catalog2} present our final catalogs, which include magnitudes (uncorrected for extinction) and their DOLPHOT uncertainty, as well as the Galactic extinction values derived for each star. Extinction-corrected photometry (denoted F606W$_0$ and F814W$_0$) is used throughout this work, and the CMDs are displayed in Figures~\ref{fig:madcash1_subaru_acs} and \ref{fig:madcash2_subaru_acs}.

\begin{table*}[tbp]
\caption{{\it HST} Photometry of MADCASH-1.} \label{tab:Catalog1}
\begin{minipage}[b]{0.95\linewidth}\centering
\begin{tabular}{ccccccccc}
\tablewidth{0pt}
\hline
\hline
Star No. & $\alpha$      &  $\delta$      & F606W     & $\delta$(F606W) & $A_{F606W}$ & F814W & $\delta$(F814W)  & $A_{F814W}$ \\
{}       & (deg J2000.0) &  (deg J2000.0) & (mag) & (mag)     & (mag)   & (mag) & (mag)     & (mag)   \\
\hline
0 & 115.70930 & 65.399063 & 19.45 & 0.01 & 0.09 & 17.53 & 0.01 & 0.05 \\
1 & 115.70916 & 65.426264 & 18.65 & 0.01 & 0.09 & 17.62 & 0.01 & 0.05 \\
2 & 115.70088 & 65.431056 & 19.14 & 0.01 & 0.09 & 17.94 & 0.01 & 0.05 \\
\hline
\end{tabular}
\caption{{\it HST} Photometry of MADCASH-2.} \label{tab:Catalog2}
\begin{tabular}{ccccccccc}
\tablewidth{0pt}
\hline
\hline
Star No. & $\alpha$      &  $\delta$      & F606W     & $\delta$(F606W) & $A_{F606W}$ & F814W & $\delta$(F814W)  & $A_{F814W}$ \\
{}       & (deg J2000.0) &  (deg J2000.0) & (mag) & (mag)     & (mag)   & (mag) & (mag)     & (mag)   \\
\hline
0 & 182.49047 & 35.456970 & 19.67 & 0.01 & 0.04 & 17.56 & 0.01 & 0.02 \\
1 & 182.50010 & 35.465970 & 18.67 & 0.01 & 0.04 & 18.21 & 0.01 & 0.02 \\
2 & 182.49053 & 35.456955 & 19.83 & 0.01 & 0.04 & 18.27 & 0.01 & 0.02 \\
\hline
\end{tabular}
   \begin{tablenotes}
      \small
      \item Notes: Star No. is our assigned number for each star. $\alpha$ and $\delta$ are the right ascension and declination, respectively. $F606W$/$F814W$ are Vega magnitudes (uncorrected for extinction), $\delta$(F606W)/$\delta$(F814W) are DOLPHOT uncertainties, and  $A_{F606W}$/$A_{F814W}$ are galactic extinction corrections in each band. 
      \item (These tables are available in their entirety in a machine-readable form in the online journal. A portion is shown here for guidance regarding its form and content.)
    \end{tablenotes}   
\end{minipage}    
\end{table*}

We derived completeness and photometric uncertainties using $\sim$100,000 artificial star tests per pointing, measured with the same photometric routines used to create the photometric catalogs. Completeness curves as a function of magnitude for each field are shown in Figure~\ref{fig:completeness}; Table~\ref{tab:obslog} shows an observation log and our estimates of the 50\% and 90\% completeness limits for each observation. The MADCASH-1 {\it HST} data are 50\% complete at F606W = 27.54~mag and F814W = 27.25~mag, while the MADCASH-2 field reaches 50\% completeness at F606W = 27.37~mag, F814W = 27.10~mag.

\begin{table}[t!]
\centering
\caption{{\it HST} observation log and field completeness.} \label{tab:obslog}
\begin{tabular}{lccccc}
\tablewidth{0pt}
\hline
\hline
Field Name    & Camera &  Filter & Exp\tablenotemark{a}  & 50\%\tablenotemark{b}  & 90\%\tablenotemark{c}  \\
{}            &  {}    &  {}     & (s)  & (mag) & (mag) \\
\hline
MADCASH~1 & ACS/WFC   &  F606W & 2475 & 27.54 & 26.65 \\
          & ACS/WFC   &  F814W & 2645 & 27.25 & 26.63 \\
\hline
MADCASH~2 & ACS/WFC   &  F606W & 2160 & 27.37 & 26.43 \\
          & ACS/WFC   &  F814W & 2200 & 27.10 & 26.51 \\
\hline
\hline
\end{tabular}
  \begin{tablenotes}
      \small
      \item $^a$ Total exposure time in seconds.
      \item $^b$ Magnitude at which the data are 50\% complete, based on artificial star tests.
      \item $^c$ Magnitude at which the data are 90\% complete.
    \end{tablenotes}   
\end{table}

\section{Structural Parameters and Physical Properties} \label{sec:dwarf_properties}

\subsection{Color-Magnitude Diagrams}
Figures~\ref{fig:madcash1_subaru_acs} (MADCASH-1) and \ref{fig:madcash2_subaru_acs} (MADCASH-2) show the {\it HST} CMDs of the two dwarfs, with a comparison to the Subaru+HSC discovery data. Both systems are clearly resolved into their constituent RGB stars in the {\it HST} data, and show old stellar populations consistent with being dwarf satellites of their putative hosts. 

\subsubsection{MADCASH-1}\label{sec:madcash1_cmd}

MADCASH-1, while visible in the image in the upper half of Figure~\ref{fig:madcash1_subaru_acs} (see also \citealt{Carlin2016}), has very few resolved RGB stars in the ground-based CMD. Furthermore, the ground-based photometry is compromised by the unresolved emission due to the numerous associated main sequence stars below the Subaru detection limit. Thus, the {\it HST} data (bottom half of Figure~\ref{fig:madcash1_subaru_acs}) are vital to (a) confirm that MADCASH-1 is a dwarf galaxy, (b) refine its distance estimate to confirm association with NGC~2403, (c) assess its stellar populations via high-quality photometry, and (d) improve our estimates of its structural parameters and luminosity. 
Given the observed lack of neutral hydrogen in MADCASH-1 \citep{Carlin2016}, we expected to find only old or intermediate-age stellar populations in the {\it HST} data. This is borne out in the lower panels of Figure~\ref{fig:madcash1_subaru_acs}, where MADCASH-1 is well resolved. Its RGB follows the most metal-poor isochrones we overplotted (in magenta), with little evidence of younger, bluer stars. In Figure~\ref{fig:madcash1_subpops}, we extract stars bluer than the RGB, as well as RGB stars filtered based on an old, metal-poor RGB, and compare their spatial positions. The small number of blue sources are not concentrated near the main body of MADCASH-1, and are thus not likely to be associated with the dwarf. We conclude that this dwarf is an old, metal-poor quenched system consisting solely of ancient stellar populations. 

\subsubsection{MADCASH-2}\label{sec:madcash2_cmd}

Figure~\ref{fig:madcash2_subaru_acs} shows the Subaru+HSC and {\it HST}+ACS images and CMDs of MADCASH-2. The system is clearly visible in the ground-based image, but crowding makes it difficult to extract reliable photometry. There are excess sources in the CMD of the upper panel relative to an equal-area background field, but the RGB is not prominent. In the {\it HST} data, the RGB is well-defined, predominantly following a metal-poor ([M/H] $= -2.0$), old (13.5~Gyr) isochrone. 

In addition to the ancient RGB stars, there is a population of stars blueward of the metal-poor RGB. 
To assess whether or not these young stars are associated with MADCASH-2, we extract stars bluer than the RGB and compare their spatial positions to those of old, metal-poor RGB stars in Figure~\ref{fig:madcash2_subpops}. There is a compact concentration of these young stars near the center of MADCASH-2 (we will show later in this work that their density profile closely matches that of the ancient RGB stars).
We conclude that these young stars are part of MADCASH-2. We further extract stars above and redward of the TRGB as candidate asymptotic giant branch (AGB) stars. Again, many of these stars (purple hexagons in Fig.~\ref{fig:madcash2_subpops}) are concentrated in MADCASH-2, suggesting that they too are associated with the dwarf. Unlike MADCASH-1 (and nearly all ultra-faint dwarf galaxies), MADCASH-2 has had fairly recent ($\sim500~$Myr ago) star formation.

To examine the young stellar population of MADCASH-2 further, we manually model its CMD using the isochrone-sampling method described in \S 3.1 of \cite{Garling2020}. In short, we specify a complex SFH by assigning relative mass ratios to PARSEC isochrones \citep{Aringer2009,Bressan2012,Chen2014,Marigo2017} of different ages and metallicities. The optimum total stellar mass for the proposed SFH is found using a minimal implementation of the likelihood function in \cite{Dolphin2002a}. Stars are then sampled from each isochrone using the \cite{Chabrier2001} log-normal initial mass function (IMF) until the limiting stellar masses are reached, at which point we mock observe the pure catalog by convolving it with the photometric completeness and error functions derived from the artificial star tests to produce a CMD comparable to what we observe in our {\it HST} imaging. We find that a SFH with 89\% of the stellar mass in a 12--13.5 Gyr, $-2.0 \leq$ [M/H] $\leq -1.5$ population, 10\% of the stellar mass in a 1.1--1.5 Gyr, [M/H]=$-1.5$ population, and $1\%$ of the stellar mass in a 400--500 Myr, [M/H]=$-1.5$ population matches the observed CMD of MADCASH-2 reasonably well. The blue loop of the young population reproduces the blue stars seen in the CMD between 25.5 $\leq$ F814W\textsubscript{0} $\leq$ 26.5 and $-0.1 \leq$ (F606W\textsubscript{0}$-$F814W\textsubscript{0}) $\leq$ 0.3, while the blue loop of the intermediate age population overlaps with the RGB given our photometric uncertainties, providing the quantity of stars fainter than F814W$_0 \gtrsim 26$ needed to match the observations. We show this model compared to MADCASH-2 and a purely old, metal-poor model population in Figure~\ref{fig:madcash2_cmd_sims}. \par

\subsection{Distances}\label{subsec:distances}

We measure distances to the two targets using the tip of the RGB (TRGB) method, which relies on the fixed luminosity of the core helium ignition stage for old stellar populations \citep[e.g.,][]{Serenelli2017}. Details on the TRGB magnitude recovery method can be found in \cite{tollerud2016}, and the adopted Bayesian code is publicly available.\footnote{https://github.com/eteq/rgbmcmr} Briefly, we use a parametrized luminosity function composed of a broken power law (including an RGB and an AGB component), which is smoothed by taking into account the photometric uncertainties derived from our artificial star tests. The color and magnitude ranges considered for MADCASH-1 and MADCASH-2 are, respectively, $22.0<$~F814W$_0<25.5$ and $0.5<$~(F606W-F814W)$_0<1.25$, and $22.0<$~F814W$_0<25.5$ and $0.25<$~(F606W-F814W)$_0<1.25$, where all magnitudes have been corrected for extinction as described above.

The derived posterior distribution for MADCASH-2 yields a distance estimate of $D_{\text{MADCASH~2}} = 3.00 ^{+0.13}_{-0.15}$~Mpc (assuming a color-dependent absolute TRGB magnitude of $M^{\text{TRGB}}_{\text{F814W}} = -4.06 + 0.2*({\rm color} - 1.23)$, where the color is the (F606W-F814W) color of the TRGB; \citealt{rizzi2007}). The TRGB magnitude and corresponding distance are reported in Table~\ref{tab:params}, and an old, metal-poor isochrone at this distance is shown in Figures~\ref{fig:madcash2_subaru_acs} and \ref{fig:madcash2_subpops}.
The dwarf's assumed host, NGC~4214, lies at a distance of 3.04 Mpc \citep[][also measured from its TRGB]{Dalcanton2009}.

The derivation of a robust TRGB distance for MADCASH-1 is complicated by the paucity of stars in the target dwarf, and by the presence of a small gap along the RGB at magnitudes F814W$\sim24$ (see Fig. \ref{fig:madcash1_subaru_acs}).\footnote{To confirm that this apparent gap in the upper RGB is simply the result of shot noise in the sparsely-populated MADCASH-1 RGB, we drew stars from synthetic stellar populations with the same number of upper RGB stars as MADCASH-1. Gaps similar to what we observe in Figure~\ref{fig:madcash1_subaru_acs} are common in the many synthetic CMDs we explored. We also note that even the synthetic CMD in the right panel of Figure~\ref{fig:madcash2_cmd_sims}, which corresponds to a more luminous dwarf, shows an underdense region near the TRGB.} Because of the sparse RGB, the Bayesian method would not converge to a reasonable solution. We thus opt to use a simpler edge-detection approach (e.g., \citealt{Lee1993}). We create a binned luminosity function (LF) of MADCASH-1 RGB stars, then smooth the LF with a zero-sum Sobel edge-detection filter. This yields a peak in the filtered LF, from which we derive a by-eye estimate of $m-M = 27.66\pm0.15$, corresponding to $D_{\text{MADCASH~1}} = 3.41 ^{+0.24}_{-0.23}$~Mpc.\footnote{As a consistency check, we applied the same technique to the MADCASH-2 luminosity function, and confirmed that we obtain a similar result (F814W$_{\rm TRGB} = 23.3\pm0.2)$ to our reported value from the Bayesian method.} This is in excellent agreement with our ground based estimate from the Subaru data of $m-M = 27.65\pm0.26$ \citep{Carlin2016}. NGC 2403, the likely host of MADCASH-1, lies at a distance of 3.09--3.20 Mpc \citep{Dalcanton2009}.

\subsection{Structural Parameters}

We derive structural parameters (including half-light radius $r_{\rm h}$, ellipticity $\epsilon$, and position angle $\theta$) for the dwarfs using the maximum-likelihood (ML) method of \citet{Martin2008}, as implemented by \citet{Sand2009}. In our analysis, we only select stars consistent with an old, metal-poor isochrone in color-magnitude space after taking into account photometric uncertainties (i.e., similar to the stars colored red in Figures~\ref{fig:madcash1_subpops} and \ref{fig:madcash2_subpops}), within our 90\% completeness limit (see Table~\ref{tab:obslog}). We inflate the uncertainty to 0.1 mag when the photometric errors are $<0.1$ mag for the purpose of selecting stars to go into our ML analysis. The resulting structural parameters are summarized in Table~\ref{tab:params}. The quoted $r_{\rm h}$ is the best-fit elliptical half-light radius along the semi-major axis. Uncertainties are determined by bootstrap resampling the data 1000 times and recalculating the structural parameters for each resample. We check our results by repeating the calculations with the same set of stars, but with a limit one magnitude brighter. The derived structural parameters using both samples of stars are consistent within the uncertainties.

\begin{figure}
\centering
\includegraphics[width=1\columnwidth, trim=0.in 0.in 0.in 0.in, clip]{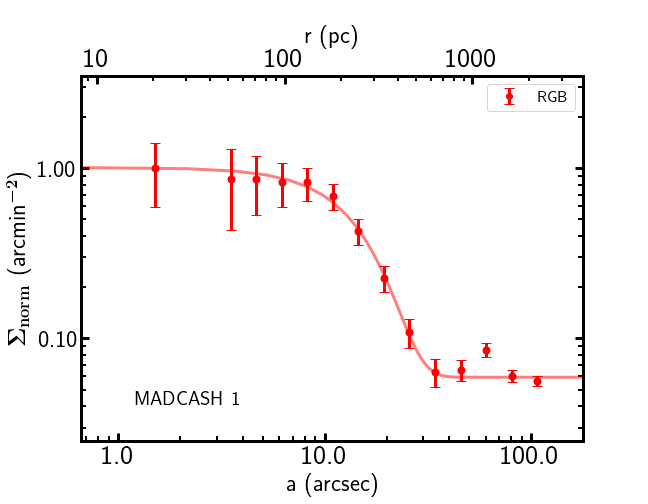}
\includegraphics[width=1\columnwidth, trim=0.in 0.in 0.in 0.in, clip]{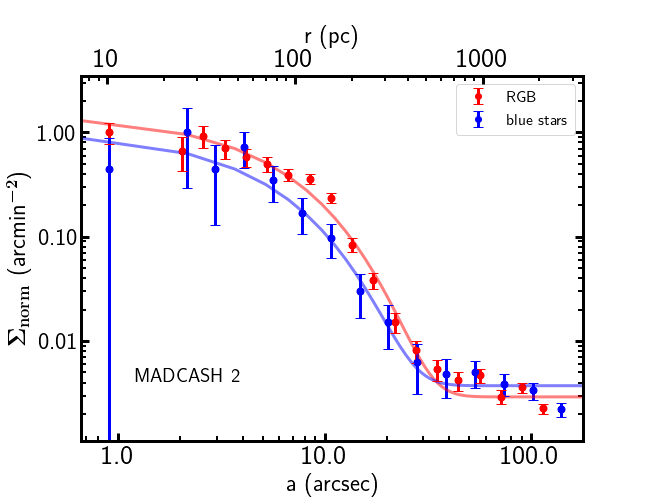}
\caption{Surface density profiles along the semi-major axes for MADCASH-1 (top) and MADCASH-2 (bottom) in stars per arcmin$^2$, but normalized so that the central density is 1.0. 
We fit exponential profiles to the surface densities, obtaining $r_{\rm h} = 13.4\pm8.4$~arcsec ($222\pm138$~pc at the dwarf's distance) for MADCASH-1.
For MADCASH-2, we separate the old, metal-poor RGB stars and the young blue-loop population. For the RGB, we obtain $r_{\rm h} = 8.1\pm1.4$~arcsec ($118\pm20$~pc at the distance of MADCASH-2), and for the blue stars the fit yields $r_{\rm h} = 7.4\pm1.3$~arcsec ($107\pm19$~pc).
The consistency of these fits suggests that the young, blue stars in MADCASH-2 follow the same spatial distribution as the old RGB population. For both systems, the fits to the binned surface density profiles agree with (but are of poorer quality than) the maximum likelihood results presented in Table~\ref{tab:params}.}
\label{fig:surfdens}
\end{figure}

In Figure~\ref{fig:surfdens}, we show one-dimensional stellar radial profiles, along with least-squares fits of exponential models to the profiles.
We use elliptical bins based on the parameters from the ML analysis, and select stars brighter than the 90\% completeness limits in each dwarf (see Section~\ref{sec:data} and Table~\ref{tab:obslog}). To facilitate comparison, we normalize each profile so that the central density is 1.0. 
From the profile fits, we obtain $r_{\rm h} = 13.4\pm8.4$~arcsec ($222\pm138$~pc) for MADCASH-1, which is consistent with the (more robust) result from the ML analysis. To look for differences between the distribution of the ancient and younger populations in MADCASH-2, we create separate density profiles for RGB and blue-loop candidates from MADCASH-2. The exponential fits to these populations yield $r_{\rm h, RGB} = 8.1\pm1.4$~arcsec and $r_{\rm h, blue loop} = 7.4\pm1.3$~arcsec.
These fits thus show that the multiple populations in MADCASH-2 are consistent with following the same spatial distributions. 
The one-dimensional representations of the exponential fits and the data are in good agreement, but we also note that parameterized models, condensed to one dimension, cannot probe a satellite's potentially complex structure.

\begin{figure}[!t]
\includegraphics[width=0.98\columnwidth, trim=0.25in 0.25in 1.0in 0.75in, clip]{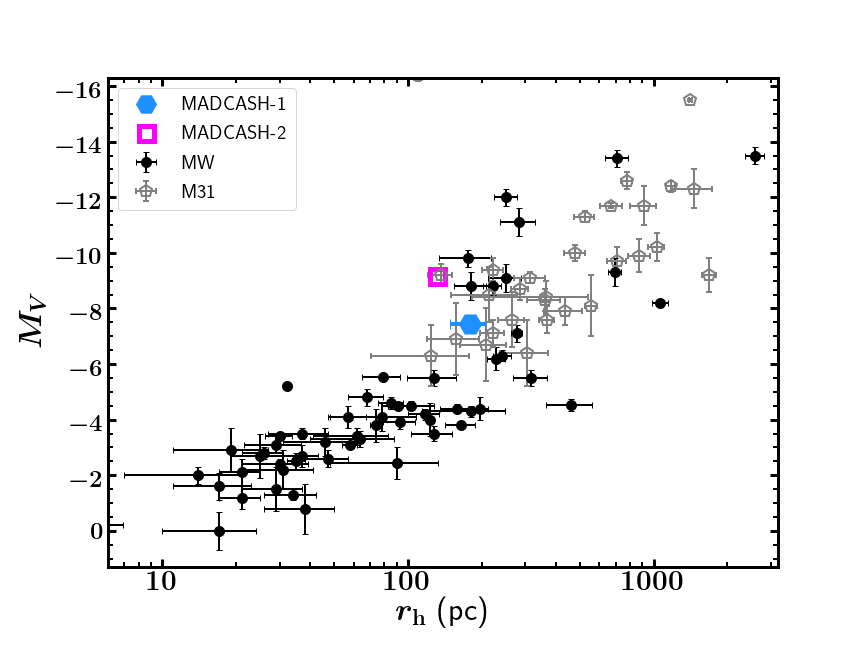}
\caption{Absolute $V$-band luminosity ($M_{{\text V}}$) vs. half-light radius ($r_{{\text h}}$) for dwarf satellites of the Milky Way (filled black circles) and M31 (open pentagons), with our results for MADCASH-1 (blue pentagon) and MADCASH-2 (open magenta square) overlaid for comparison. The MADCASH dwarfs are generally consistent with MW/M31 satellites in this size-luminosity plane. While MADCASH-2 is smaller than all of the classical dSphs of the MW and M31, it is adjacent to the locus occupied by LG dSphs, so this may not be a significant difference. Data for MW/M31 dwarfs are predominantly from \citet[][see references therein]{McConnachie2012}, with additional dwarfs (or updated measurements) from \citet{Drlica-Wagner2015,Kim2015, Kim2015a,Laevens2015,Laevens2015a, Martin2015, Crnojevic2016, Drlica-Wagner2016,Homma2016,  Torrealba2016,Torrealba2016Crater2, Carlin2017AJ, Homma2018,Burcin2018, Torrealba2018,Homma2019, Mau2019,Cerny2020, Mau2020}.
}
\label{fig:mv_rh}
\end{figure}

\subsection{Luminosities} \label{subsection:luminosities}
We derive absolute magnitudes for our objects by using the same procedure as in \citet{Burcin2018}, 
as was first described in \citet{Martin2008}. First, we build a well-populated CMD (of $\sim$ 20,000 stars), including our completeness and photometric uncertainties, by using a PARSEC isochrone with a metallicity of [M/H]$=-2.0$ and age 13.5~Gyr and its associated luminosity function assuming a \citet{Chabrier2001} IMF. Then, we randomly select the same number of stars from this artificial CMD as was found from our exponential profile fits (over the same magnitude range as used for the ML analysis). 

We sum the flux of these stars, and extrapolate the flux of unaccounted stars (i.e., those fainter than the detection limit) using the adopted luminosity function. We calculate 1000 realizations in this way, and take the mean as our absolute magnitude and its standard deviation as the uncertainty. To account for the uncertainty on the number of stars, we repeat this operation 100 times, varying the number of stars within their uncertainties, and use the offset from the best-fit value as the associated uncertainty. These error 
terms and distance modulus uncertainty are then added in quadrature to produce our final uncertainty on the absolute magnitude.

The results of our luminosity estimates for the two dwarfs are reported in Table~\ref{tab:params}. We find $M_{\text{V}} = -7.81\pm0.18$ for MADCASH-1. This much more robust measurement is consistent with our estimate of $M_{\text{V}} = -7.7\pm0.7$ from the ground-based Subaru data \citep{Carlin2016}. Our improved measurement confirms the status of MADCASH-1 as the faintest known dwarf companion of an isolated MC-mass host. For MADCASH-2, we find $M_{\text{V}} = -9.15\pm0.12$, placing this system's luminosity near the faint end of those of the ``classical'' dSphs in the MW and M31. Assuming an average $V-$band mass-to-light of $M_{\text{star}}/L_{\text{V}} = 1.6$ \citep{Woo2008} appropriate for old stellar populations, the measured luminosities correspond to stellar masses of $M_{\text{star}} = (1.8\pm0.3) \times 10^5~M_\odot$ and $(6.3\pm0.6) \times 10^5~M_\odot$ for MADCASH-1 and MADCASH-2, respectively. Figure~\ref{fig:mv_rh} shows these companions of MC-mass hosts in context with MW and M31 dwarfs in the size-luminosity plane. The MADCASH dwarfs occupy positions consistent with those of Local Group dSphs; their half-light radii are on the small end for their luminosities, but they are not significant outliers relative to LG systems.

These total luminosities were estimated based solely on the old, metal-poor population. As noted in Section~\ref{sec:madcash2_cmd} and Figure~\ref{fig:madcash2_cmd_sims}, MADCASH-2 apparently contains a small number of relatively young, metal-poor stars. To estimate the total mass contained in this young population, we perform a simple exercise. We create artificial stellar populations with [M/H]$=-2.0$ and age 1~Gyr based on PARSEC isochrones \citep{Aringer2009,Bressan2012,Chen2014,Marigo2017} populated by a Chabrier log-normal IMF \citep{Chabrier2001} and shifted to the distance of MADCASH-2. We then select mock stars at random, summing the stellar mass of the selected stars, until the total number of artificial stars within the same blue box used to select the blue loop stars in Figures~\ref{fig:madcash2_subpops} and \ref{fig:madcash2_cmd_sims} is equal to the number of candidate blue loop stars in MADCASH-2 (a total of 22$\pm$5 within 2~$r_{\text{h}}$). This yields the total mass in stars that corresponds to a sample containing the same number of blue loop stars as the young population in MADCASH-2. We perform this exercise 1000 times, tabulating the total stellar masses, and find a mean stellar mass of $2760\pm550~M_\odot$ in the young population.\footnote{The same process executed instead with a Salpeter IMF yields a stellar mass of $2460\pm490~M_\odot$. This mass estimate is slightly lower than the one based on a Chabrier IMF, but consistent within the uncertainties.} This population thus makes up just less than 1\% of the total stellar mass of MADCASH~2, consistent with the estimate from the CMD modelling in \S \ref{sec:madcash2_cmd}. Finally, we note that the ``blue loop'' of the intermediate-age population is at fainter magnitudes than the younger blue loop stars, and thus does not contaminate the selection used for this calculation.

\subsection{Metallicities}
Finally, we estimate metallicities by creating a grid of old (10 Gyr) isochrones at values of $-2.5 < {\rm [M/H]} < 0.0$, spaced at 0.1-dex intervals.\footnote{Although we have adopted PARSEC isochrones throughout this work, for this exercise we used Dartmouth isochrones \citep{Dotter2008} because they extend to lower metallicities than the PARSEC grid. This should make little difference to this coarse estimate of metallicity. }
Fixing each isochrone at our derived TRGB distance, we sum the distances (in the CMD) of all RGB stars from the selected isochrone (as to only include the dominant ancient, metal-poor population). The best-fitting isochrone is the one for which this summed residual is minimized. For each dwarf, we report a best-fit isochrone metallicity of [M/H] = $-2.0$. However, there is little separation between isochrones at [M/H] = $-2.5$ and [M/H] = $-2.0$, so that we are unable to distinguish between isochrones at the metal-poor end. Thus, it is more accurate to report [M/H]$ \leq -2.0$ for the old RGB populations in both MADCASH-1 and MADCASH-2. 

\subsection{Search for globular clusters}

We also searched the {\it HST} images for the possible presence of globular star clusters associated with the dwarfs. Galaxies of these masses sometimes host such clusters, which can have important implications for the formation and structure of their halos \citep[e.g., ][]{Crnojevic2016, Cusano2016, Amorisco2017,Caldwell2017,Li2017, Contenta2018}. Both sets of images contain a number of slightly extended, round sources that would have inferred masses consistent with relatively low-mass ($\lesssim 10^5 M_{\odot}$) globular clusters. However, the properties of these sources are also consistent with their being background galaxies, and none of them are located within $\sim 5 r_h$ of either dwarf. In the absence of spectroscopy, we cannot rule out the possibility that one of these might indeed be a true, distant star cluster, but we see no clear evidence for obvious clusters associated with either dwarf.

\subsection{Neutral hydrogen in MADCASH-2}\label{sec:HI}
    
We obtained position-switched HI observations of MADCASH-2 using the Robert C. Byrd Green Bank Telescope\footnote{The Green Bank Observatory is a facility of the National Science Foundation operated under cooperative agreement by Associated Universities, Inc.}(AGBT19B-144; PI: Karunakaran) with VEGAS in Mode 7 (bandwidth = 11.72 MHz). We searched the full velocity range of $-1100~{\rm km ~s}^{-1} \leq V_{Hel} \leq -150~{\rm km ~s}^{-1}$ and $50~{\rm km ~s}^{-1} \leq V_{Hel} \leq 1300~{\rm km ~s}^{-1}$ for any HI emission associated with MADCASH-2 and found none. The host, NGC 4214, has a systemic velocity of $291~{\rm km ~s}^{-1}$ \citep{Walter2008}. The GBT spectrum has an rms noise of $\sigma=0.37~{\rm mJy}$ in the aforementioned velocity range at a velocity resolution of $15~{\rm km ~s}^{-1}$. Using the measured distance, V-band luminosity, and an assumed velocity distribution of $15~{\rm km ~s}^{-1}$ we place stringent $5\sigma$, single-channel upper limits of $M_{\rm HI} < 4.8\times10^4 M_\odot$ and $M_{HI}/L_V < 0.08 M_\odot/L_\odot$. This result is commensurate with MADCASH-1 \citep{Carlin2016} and other gas-poor dwarf spheroidals in the Local Volume \citep{Grcevich2009,Spekkens2014,Karunakaran2020}, but is strange given the recent star formation suggested by MADCASH-2's CMD (Figure \ref{fig:madcash2_cmd_sims}). 
In Section~\ref{sec:mc2}, we will discuss possible explanations for the presence of young stars but lack of H\textsc{I} in MADCASH-2.

\section{Discussion} \label{sec:discussion}

The {\it HST} follow-up observations we have presented provide confirmation that MADCASH-1 and MADCASH-2 are indeed low-mass dSphs at the same distances as their putative hosts. In this section we provide context about the place of these two dwarfs within their hosts' satellite systems, and within our knowledge of dwarf satellite properties more generally. What emerges is a scenario in which shaping of satellite properties by environmental processes is more important around MC-mass hosts than previously thought.

\subsection{MADCASH-1 and the NGC~2403 system}\label{sec:mc1}

\citet{Dooley2017b} predicted that NGC~2403 should have $\sim 2-6$ dwarf companions with stellar masses $> 10^5 M_\odot$ ($M_V \lesssim -7$) within 100~kpc of its center, with the scatter in predictions depending in part on the adopted abundance matching prescription, and also on factors such as infall time into the host halo and the time at which reionization occurred (as detailed in \citealt{Dooley2017a}). With the known massive companion DDO~44 shown to be tidally disrupting, and thus originally even more luminous than its present total (dwarf plus stream) luminosity of $M_{\rm V} \sim -12.9$, by \citealt{Carlin2019}, MADCASH-1 brings the total number of known NGC~2403 satellites to two. We have recently completed observations of the entire $\gtrsim100$~kpc radius region around NGC~2403, and a publication with robust statistical limits on the number of satellites in this area is in preparation. At present we can tentatively note that there appear to be only two dwarf companions brighter than $M_{\rm V} = -7$ around NGC~2403, which falls at the low end of the predictions from \citet{Dooley2017b}.

Our measured luminosity for MADCASH-1 ($M_{\rm V} = -7.81\pm+0.18$; Table~\ref{tab:params}) places it right at the typical threshold for ultra-faint dwarfs ($M_{\rm V} < -7.7$; \citealt{Simon2019}). Regardless of whether we call it a UFD, its properties are typical of Local Group dwarfs of similar luminosity. For example, its half-light radius ($r_{\rm h} = 178^{+30}_{-28}$~pc) is consistent with sizes of Milky Way and M31 dwarfs at similar luminosities (Fig.~\ref{fig:mv_rh}). It is also metal-poor, as expected based on the stellar mass-metallicity relation for dwarf galaxies (e.g., \citealt{Kirby2013a}). From our {\it HST} CMD (Figure~\ref{fig:madcash1_subaru_acs}), it is apparent that MADCASH-1 hosts only an old, metal-poor RGB, and is thus similar to UFDs in the MW, which typically consist of only ancient, very metal-poor stellar populations. As shown in \citet{Carlin2016}, MADCASH-1 has only $< 7.1\times10^4~M_\odot$ of neutral hydrogen. The absence of a significant gas reservoir is also consistent with its lack of young stars, and is typical of galaxies of this stellar mass observed or predicted to be in dense environments \citep{applebaum2020,akins2020,digby2019,Karunakaran2020,rey2020}.

\subsection{MADCASH-2 and the NGC~4214 system}\label{sec:mc2}

Our measured luminosity ($M_{\rm V} = -9.15\pm0.12$) for MADCASH-2 places it near the faint end of MW ``classical'' dwarfs. However, most of these dwarfs show very little star formation in the last 3 Gyr, as shown in Figures 7 and 9 of \cite{Weisz2014a}, while our SFH modelling suggests MADCASH-2 formed about $11\%$ of its stellar mass in the last 1.5 Gyr. Two dwarfs of comparable luminosity and recent star formation are Leo~T (MW satellite with $D=407$ kpc, $M_{\text{V}}=-8$; \citealt{Weisz2012}) and Antlia B (NGC 3109 satellite with $D=1.35$ Mpc, $M_{\text{V}}=-9.4$; \citealt{Sand2015,Hargis2020}). Both dwarfs appear to have formed $\sim20\%$ of their stellar mass in the last 3 Gyr, but Leo T's SFH is more constant over this length of time compared to Antlia B; Leo T formed perhaps $10\%$ of its stellar mass in the last 1.5 Gyr, while Antlia B appears to have only formed $1\%$ in this time, compared to the estimated $11\%$ of MADCASH-2. Both Leo T and Antlia B are gas-rich, but Antlia B has 
a negligible present-day star formation rate, having formed its last stars $\sim300$ Myr ago, while Leo T has been forming stars up until as recently as 25 Myr ago. In contrast, MADCASH-2 is gas-poor, despite the CMD suggesting a star formation episode as recently as 400--500 Myr ago. Interestingly, all three dwarf galaxies are candidates for ``reignited" dwarf galaxies -- galaxies that ceased forming stars shortly after reionization, but that retained and accreted H\textsc{i} at late times to induce a recent epoch of star formation \citep{jeon2017,wright2019,rey2020}.\par

What is particularly interesting about comparing these three galaxies is the difference in their environments. Leo T is more than 400 kpc from the Milky Way and gas-rich, indicating there is a good chance it is on its first infall into the system. Given current measured distances, Antlia~B is $\sim100$ kpc from NGC~3109, making it part of the dwarf association encompassing Sextans A and B, Antlia, and Leo P. Meanwhile, MADCASH-2 is most likely a dwarf satellite of NGC~4214, an LMC-analog with only one other known dwarf galaxy (DDO~113) that is considerably brighter and ceased forming stars about 1 Gyr ago ($M_{\text{V}}=-12.2$; \citealt{Garling2020,Weisz2011}). As shown in \cite{Garling2020}, the cessation of cold gas inflows upon entering the halo of NGC~4214 (i.e., strangulation) is the most likely cause for the quenching of DDO~113. 

We reiterate that the luminosity and stellar mass presented in Table \ref{tab:params} are based solely on comparisons to old (13.5~Gyr) populations. From the models that include young and intermediate stellar populations as presented in Figure \ref{fig:madcash2_cmd_sims}, we infer an absolute magnitude $M_V=-8.95\pm0.2$, stellar mass $M_{\rm star}= (2.95 \pm 0.5) \times10^5 M_{\odot}$, and stellar mass-to-light ratio $\Upsilon_{\rm star}=0.9 \pm 0.2$ for MADCASH-2. This absolute magnitude is consistent with the measurement made under the assumption of a purely old population in \S \ref{subsection:luminosities} of $M_V=-9.15 \pm0.12$. However, the inferred stellar mass from the complex stellar population model is about a factor of two lower than the stellar mass estimated from the purely old model, due to the assumption therein of a higher stellar mass-to-light ratio ($\Upsilon_{\rm star}=1.6$) than we derived for the complex stellar population model. We note that these models are simple and should be viewed predominantly as supporting evidence for the presence of younger stellar populations in MADCASH-2, but we include these measurements based on the models to highlight the differences in inferred quantities that ignoring the younger stellar populations might create. \par

Other than the young stellar population, the properties of MADCASH-2 are typical of LG dSphs of similar luminosities. Its size is on the small side of, but consistent with, the locus of LG dSphs in the size-luminosity plane (Fig.~\ref{fig:mv_rh}), and it consists of predominantly metal-poor ([M/H]$\sim-$~2.0) stars, as is usual for faint dwarfs in the LG (e.g., \citealt{Simon2019}). 

MADCASH~2 is located $\sim70$~kpc in projection from NGC~4214. We estimate the minimum velocity of MADCASH~2 relative to NGC~4214 assuming that it must be at least 70~kpc from NGC~4214 at present, and that the most recent star formation was triggered by its pericentric passage between $\sim0.5-1.0$~Gyr ago. By this simple argument, MADCASH~2 would have to be traveling at $\gtrsim140$~km~s$^{-1}$ (on average) relative to NGC~4214 to have moved 70~kpc in 500~Myr. Alternatively, MADCASH-2 may have recently fallen into the NGC~4214 halo for the first time, though we consider it more likely that the shock required to ignite its recent burst of star formation was imparted by interaction with its more massive host NGC~4214.

\subsection{Broader context} \label{subsection:quenching}

In this section, we view MADCASH-1, MADCASH-2, and their systems through the lenses of $\Lambda$CDM predictions for low-mass host satellite systems and  environmental quenching of star formation.  Because the hosts of our two satellite candidates -- NGC 2403 and NGC 4214 -- are considerably lower-mass systems than the MW scale that receives outsized attention, we have an opportunity to explore whether $\Lambda$CDM predictions of hierarchical structure formation hold for lower density environments, and investigate environmental processing of dwarf galaxies in lower density (but not field) environments.

Although our studies of the virial volumes of NGC 2403 and NGC 4214 are not yet complete, we may assess whether the known satellites -- DDO 44 and MADCASH-1 for NGC 2403, and DDO 113 and MADCASH-2 for NGC 4214 -- are consistent with $\Lambda$CDM expectations. In the previous subsections, we compared the number of satellites we found with $M_{\rm star} \gtrsim 10^5 M_\odot$ with the semi-empirical predictions from \citet{Dooley2017a}, finding that the two satellites we found in each system was consistent with, but at the low end of, the predictions.   We may also compare with other work, namely the semi-analytic predictions for and observations of the satellite luminosity functions for galaxies outside the Local Volume in the Sloan Digital Sky Survey \citep{blanton2005,abazajian2009}.  \citet{Sales2013} find that the $\Lambda$CDM-based semi-analytic model of \citet{guo2011} is well-matched to their own measurement of satellite luminosity functions based on DR7 of the Sloan Digital Sky Survey.  Although their data set is significantly shallower than our deep MADCASH data, we can extrapolate their predictions to fainter magnitudes assuming that the faint-end slope of the satellite luminosity function remains a power law.  Based on \citet{Sales2013}, we expect to find approximately $1-7$ satellites brighter than MADCASH-1 in each system, consistent with the predictions of \citet{Dooley2017a} and our observations of the NGC 2403 and NGC 4214 systems.

We turn next to the star-formation properties of the four dwarf galaxy satellites of these systems.  None of the satellites are actively star-forming, and none show the presence of H$\textsc{i}$.  The mechanisms and timescales for the quenching of dwarf galaxies are hotly debated topics.  Much of the attention in recent years has focused on the environments of the MW and MW analogs.  Famously, the satellites of the MW and Andromeda are gas-poor and quenched \citep[e.g.,][]{Grcevich2009,Spekkens2014}, and mechanisms including tidal heating/stripping, ram-pressure stripping, strangulation, and internal quenching have been proposed to act in concert to end star formation and blow out the remaining H\textsc{i} for classical dwarf galaxies \citep[e.g.,][]{mayer2006,nichols2011,Gatto2013,slater2014,fillingham2018,fillingham2019,digby2019,simons2020,akins2020}.  It is still unknown what the relative importance of these effects are for understanding the full Local Group classical satellite population.  For small galaxies ($M_{\rm star} \lesssim 10^5 M_\odot$), the main observational sample comes from the MW, and those satellites appear quenched by reionization \citep[e.g.,][]{brown2014,Weisz2014,rey2019,rodriguezwimberly2019}.  

However, for other MW-mass hosts, many satellites are actively star-forming \citep[e.g.,][]{Geha2017,Mao2020,dickey2020}.  Observationally, \citet{Karunakaran2020} find a clear distinction for MW-mass hosts: satellites fainter than $M_V > -12$ \citep[corresponding approximately to the SAGA survey's limiting magnitude;][]{Mao2020} are generally quenched, but brighter satellites are heterogeneous in their star-forming properties.  This limiting magnitude is similar to the absolute magnitudes of DDO 113 and DDO 44 (before it was stripped).  Thus, the lack of ongoing star formation and H\textsc{i} in these galaxies and in MADCASH-1 and MADCASH-2 would be unsurprising if they were satellites of MW-mass hosts.

There is little theoretical study of the environmental effects of low-mass hosts on their satellites, but we naively expect low-mass hosts to be gentler on their satellites than a MW-mass host might be.  The shallower potential well implies a weaker tidal field; low-mass hosts like ours are not expected to have hot coronae \citep[e.g.,][]{birnboim2003,brooks2009,keres2009}, thus weakening effects like ram-pressure stripping and strangulation.  However, because mass-loading in outflows increases for decreasing stellar mass, low-mass galaxies can have massive, complex warm and cool circumgalactic media \citep[CGM;][]{bordoloi2014,muratov2015,lu2015,christensen2016,johnson2017,hafen2019}.  \citet{Garling2020} argue that this CGM may play a role in disconnecting dwarf satellites from their source of accreted gas, and thus lead to strangulation or even ram-pressure stripping.  The magnitude of these effects remains unclear.

Curiously, the vast majority of known Local Volume satellites of the LMC- and SMC-mass hosts that are the focus of the MADCASH survey are no longer star-forming, and many show clear signs of environmental quenching. There are but stringent upper limits on the H\textsc{i} content of the four known dwarf satellites of NGC 2403 and NGC 4214.  Even for MADCASH-1, a possible UFD, a lack of H\textsc{i} would be surprising if it were in the field \citep[e.g.,][]{jeon2017,rey2020}.  For MADCASH-2, DDO 44, and DDO 113, star formation almost certainly quenched after infall onto their hosts, suggesting that the tidal field and/or CGM play a significant role in driving gas out of, and ending star formation in, these satellites.  Moreover, the handful of dwarf satellites of other nearby LMC analogs also appear quenched or well on their way.  Donatiello I around NGC~404 appears to be an ancient Draco-like satellite \citep{Martinez-Delgado2018}.  NGC 3109 is closer to the SMC in mass rather than the LMC, yet its two likely satellites, Antlia and Antlia~B, are either only barely forming stars (Antlia) or not at all (Antlia~B).  Unlike the other satellites discussed here, these two hold onto significant H\textsc{i}, but that of Antlia is highly disturbed \citep{Ott2012,Hargis2020}.  Our work on the H\textsc{i} structure of Antlia~B is forthcoming.  In short,  satellites of low-mass hosts experience significant environmental processing, the precise origin of which is unknown.

With imaging from MADCASH of the full virial volumes of multiple LMC-scale hosts, we will assemble dwarf samples with well-characterized selection functions that will allow us to test more complex models and make more definitive statements about the environmental effects LMC-scale hosts have on their satellite populations. Overall, the apparent quenching of star formation in MADCASH-2, Antlia B, DDO 44, and DDO 113 suggests that even LMC-scale hosts can quench star formation in satellites, and encourages continued observational efforts to discover and characterize dwarfs of low-mass hosts, along with new theoretical work to determine the mechanisms responsible for their quenching.

\begin{deluxetable}{ccc}
\tablecaption{{\it HST}-derived properties of MADCASH dwarfs \label{tab:params}}
\tabletypesize{\scriptsize}
\tablehead{\colhead{Parameter} & \colhead{MADCASH-1\tablenotemark{a}} & \colhead{MADCASH-2\tablenotemark{b}}}
\startdata
Likely host               & NGC 2403                  & NGC 4214 \\
{RA} (hh:mm:ss)      & $07:42:39.40\pm2.50''$    & $12:10:06.74\pm0.50''$\\
{Decl} (dd:mm:ss)    & $+65:25:00.01\pm1.40''$  & $+35:26:34.58\pm0.40''$\\
{(RA, Decl)} (deg) & (115.6642, 65.4167) & (182.5281, 35.4430)\\
F814W$_{\rm TRGB}$ & $23.55\pm0.15$            & $23.29^{+0.09}_{-0.11}$\\
$m-M$ (mag)\tablenotemark{c} & $27.66 \pm 0.15$ & $27.39^{+0.09}_{-0.11}$\\
$D$ (Mpc)          & $3.41^{+0.24}_{-0.23}$    & $3.00^{+0.13}_{-0.15}$ \\
${\rm [M/H]}_{\rm iso}$\tablenotemark{d} & $-2.0$ & $-2.0$\\
$M_{V}$ (mag)      & $-7.81\pm0.18$            & $-9.15\pm0.12$\\
$r_{\text{h}}$ (arcsec) & $10.8 \pm 1.0$           & $9.0 \pm 0.5$\\
$r_{\text{h}}$ (pc)   &  $178.5^{+30.3}_{-27.5}$  & $130.9^{+13.3}_{-13.5}$\\
$\epsilon$         & $0.25\pm0.11$             & $0.19\pm0.05$\\
$\theta$ (deg)    & $0^\circ \pm 19^\circ $   & $76^\circ \pm 11^\circ $ \\
$\mu_{\text{V,eff}}$ (mag arcsec$^{-2}$)\tablenotemark{e} & $26.7 \pm 0.4$   & $24.8 \pm 0.3$\\
$M_{\rm star} (M_\odot$)\tablenotemark{f} & $(1.8\pm0.3) \times10^5$   & $(6.3\pm0.6) \times10^5$ \\
$M_{\rm HI}/L_{V} (M_\odot/L_\odot)$ & $<0.39$ & $<0.08$ \\
$M_{\rm HI} (M_\odot)$ & $<7.1\times10^4$      & $<4.8\times10^4$ \\
\enddata
\tablenotetext{a}{Following IAU naming conventions, this dwarf was dubbed MADCASH J074238+652501-dw in \citet{Carlin2016}.}
\tablenotetext{b}{Following the naming convention used for the first MADCASH dwarf, this dwarf would be designated MADCASH J121007+352635-dw.}
\tablenotetext{c}{Assuming $M^{\rm TRGB}_{F814W} = -4.06 + 0.2*({\rm color} - 1.23)$ \citep{rizzi2007}, where $color$ is the (F606W-F814W) color of the TRGB.}
\tablenotetext{d}{Estimated using isochrones of age 10 Gyr; the reported value is just an approximation, see text for details.}
tablenotemark{e}{Mean surface brightness within the (elliptical) half-light radius.}
\tablenotetext{f}{Assuming $M_{\rm star}/L_{\rm V} = 1.6$, as is typical for dSphs \citep{Woo2008}.}
\end{deluxetable}

\section{Conclusions}

In this work, we have presented Subaru+HSC and {\it HST}+ACS observations of two dwarf spheroidal galaxies that are likely satellites of host galaxies with stellar masses similar to that of the Large Magellanic Cloud. The first of these, MADCASH-1, is the first ultra-faint dwarf galaxy found orbiting a $\sim$LMC-mass host. The second, MADCASH-2, looks similar to typical faint dSphs, except that it shows evidence of multiple recent ($\sim400$~Myr and 1.5~Gyr ago) star formation episodes. Both galaxies were discovered in deep Subaru imaging data -- we were able to measure the structural properties of MADCASH-1 based on the ground-based data \citep{Carlin2016}, but with large uncertainties, while MADCASH-2 is too compact to extract reliable measurements of its individual stars. The {\it HST} results presented here represent the first confirmation of MADCASH-2 as a dSph likely associated with NGC~4214, and a more robust confirmation of MADCASH-1's association with NGC~2403, while also enabling detailed derivation of their structural parameters.

{\it HST} photometry reaching $>3.5$~mag below the TRGB reveals that MADCASH-1 (officially designated MADCASH J074238+652501-dw) consists solely of old ($>10$~Gyr), metal-poor ([M/H] $\sim -2.0$) stellar populations. The superior resolution of {\it HST} allows us to measure the detailed properties of MADCASH-1. We find a total luminosity of $M_{\text{V}} = -7.81\pm0.18$, which corresponds to a stellar mass of $\sim1.8\times10^5~M_\odot$. Our estimate of the TRGB distance to MADCASH-1 ($D = 3.41^{+0.24}_{-0.23}$~Mpc) is consistent with the system being a satellite of NGC~2403 ($D \sim 3.09-3.20$~Mpc; e.g.,  \citealt{Dalcanton2009, Jacobs2009, Radburn-Smith2011}), a galaxy with roughly 2-3 times the stellar mass of the LMC ($M_{\rm star} \sim 3.7-5.1\times10^9 M_{\odot}$; \citealt{DeBlok2008,Leroy2019}). The structural parameters (e.g., $r_{\text{h}}$, ellipticity, position angle, mean surface brightness) we derive for MADCASH-1 are typical of UFDs with similar luminosities. 

We also present the discovery of a dSph that is likely associated with NGC~4214, a nearby galaxy with stellar mass similar to that of the LMC ($M_{\rm star} \sim 3.29\times10^9~M_\odot$; \citealt{Weisz2011}). The new dwarf, MADCASH-2 (officially designated MADCASH J121007+352635-dw), was discovered in a visual search of deep Subaru+HSC images. The compactness of the system made it difficult to extract photometry of individual stars, so we proposed for {\it HST}+ACS observations. MADCASH-2 is well-resolved in {\it HST} images, revealing a well-populated old ($>10$~Gyr), metal-poor ([M/H] $\sim -2.0$) RGB. Our derived distance for MADCASH-2 ($D = 3.00^{+0.13}_{-0.15}$~Mpc) is consistent with an association between the dwarf and NGC~4214 ($D = 3.04$~Mpc; \citealt{Dalcanton2009}). The system's luminosity, $M_{\text{V}} = -9.15\pm0.12$, places it at the faint end of the classical dwarfs. The structural properties and metallicity of MADCASH-2 are consistent with typical classical dwarfs of similar luminosity. 

The majority of the stars resolved in the {\it HST} observations of MADCASH-2 are consistent with an old, metal-poor population, but there is a substantial population of blue loop stars that we associate with $\sim400$ Myr and 1.5 Gyr populations. Their formation may have been triggered by a recent pericentric passage about NGC~4214, and perhaps even the first pericenter of a recently-infallen MADCASH-2. With a dedicated search using the GBT, we find no evidence of neutral hydrogen in MADCASH-2, so the recent star formation episode must have either exhausted or expelled its gas, or else the gas was lost due to tidal forces shortly after the interaction that precipitated the brief star formation event. The other, brighter dwarf satellite of NGC 4214 (DDO 113; $M_{\text{V}}=-12.2$) is also found to be quenched \citep{Weisz2011}, likely due to strangulation \citep{Garling2020}. Discovery that MADCASH-2 formed stars recently but is currently quiescent with no detectable neutral hydrogen lends further support to theories suggesting that LMC-scale hosts have an environmental impact on their satellite populations despite their low masses, but further investigation is needed to determine what mechanisms are responsible.\par

The results reported here characterize the first dwarf satellites of Magellanic analogs discovered in our MADCASH survey. Ultimately, these systems will be part of a statistically complete census of the outskirts of Local Volume MC analogs, with which we will derive satellite luminosity functions and characterize the properties of dwarfs around many host systems in a variety of environments.

\acknowledgments

We thank the referee for comments that helped clarify this work. JLC acknowledges support from HST grant HST-GO-15228
and National Science Foundation (NSF) grant AST-1816196. 
BMP is supported by an NSF Astronomy and Astrophysics Postdoctoral Fellowship under award AST-2001663.
DC and SL are supported by NSF grant AST-1814208.
Research by DJS is supported by NSF grants AST-1821967 and AST-1813708. CTG and AHGP are supported by NSF grant AST-1813628. JS acknowledges support from the Packard Foundation. 
The work of authors JLC, DJS, and JRH was performed in part at the Aspen Center for Physics, which is supported by NSF grant PHY-1607611.
AJR was supported by NSF grant AST-1616710, and as a Research Corporation for Science Advancement Cottrell Scholar. KS acknowledges support from the Natural Sciences and Engineering Research Council of Canada (NSERC).

Based on observations made with the NASA/ESA Hubble Space Telescope, obtained at the Space Telescope Science Institute, which is operated by the Association of Universities for Research in Astronomy, Inc., under NASA contract NAS5-26555. These observations are associated with program \#15228. Support for program \#15228 was provided by NASA through a grant from the Space Telescope Science Institute, which is operated by the Association of Universities for Research in Astronomy, Inc., under NASA contract NAS5-26555.

This research has made use of NASA's Astrophysics Data System, and \texttt{Astropy}, a community-developed core Python package for Astronomy \citep{Price-Whelan2018b}. This research has made use of ``Aladin sky atlas'' developed at CDS, Strasbourg Observatory, France.

%

\vspace{5mm}
\facilities{HST, Subaru+HSC, GBT}


\software{\texttt{astropy} \citep{TheAstropyCollaboration2013,Price-Whelan2018b}, \texttt{DOLPHOT} \citep{Dolphin2002}, \texttt{Matplotlib} \citep{Hunter2007}, \texttt{NumPy} \citep{VanderWalt2011}, \texttt{Topcat} \citep{Taylor2005}.}

\bibliographystyle{aasjournal}

\end{document}